\documentclass[preprint, 3p, times]{elsarticle}

\usepackage{amsthm}

\usepackage{scalerel}
\usepackage{amssymb}
\usepackage[utf8]{inputenc}  
\usepackage{url}
\usepackage{graphicx}
\usepackage{caption}
\usepackage{subcaption}
\usepackage[usenames,dvipsnames]{color}
\usepackage{xcolor}
\usepackage{alltt}
\usepackage{multicol}
\usepackage{amsmath}
\usepackage{amsfonts}
\usepackage{listings}
\usepackage{mathtools}
\usepackage{syntax}
\usepackage{ragged2e}
\usepackage{fancybox}
\usepackage{tikz}
\usepackage{xstring}
\usepackage{listings}
\usepackage[ruled,vlined,linesnumbered]{algorithm2e}

\SetCommentSty{mycommfont}

\usetikzlibrary{shapes,positioning,decorations.pathmorphing}
\lstdefinelanguage{Hrebeca}{
    morekeywords={
        softwareclass, physicalclass, changemode, mode, inv, msgsrv, guard, delay, 
        statevars, int, float, real, knownrebecs, if, else, self, main, Wire, CAN, 
        bool, reactiveclass, true, false, setmode, after
    },
    otherkeywords={=>,<-,<\%,<:,>:,\#,@},
    sensitive=true,
    morecomment=[l]{//},
    morecomment=[n]{/*}{*/},
    morestring=[b]",
    morestring=[b]',
    morestring=[b]""",
    literate={'}{'}1,
}
\newenvironment{Proof}{\noindent\textit{Proof.~}}{\hfill $\blacksquare$}
\newenvironment{ProofSketch}{\noindent\textit{Proof sketch.~}}{\hfill $\blacksquare$}
\newtheorem{definition}{Definition}
\newtheorem{example}[definition]{Example}
\newtheorem{lemma}[definition]{Lemma}
\newtheorem{theorem}[definition]{Theorem}

\newcommand{\Rule}[2]{                                  
	\frac{\raisebox{.7ex}{\normalsize{$#1$}}}
	{\raisebox{-1.0ex}{\normalsize{$#2$}}}}

\newcommand{\generalIntv}[4]{
	\IfEqCase{#1}{%
		{i}{
			\IfEqCase{#4}{%
				{i}{
					[ {#2} , {#3} ]
				}
				{n}{
					[ {#2} , {#3} )
				}
				{g}{
					[ {#2} , {#3} \succ
				}
			}
		}
		{n}{
			\IfEqCase{#4}{%
				{i}{
					( {#2} , {#3} ]
				}
				{n}{
					( {#2} , {#3} )
				}
				{g}{
					( {#2} , {#3} \succ
				}
			}
		}
		{g}{
			\IfEqCase{#4}{%
				{i}{
					\prec {#2} , {#3} ]
				}
				{n}{
					\prec {#2} , {#3} )
				}
				{g}{
					\prec {#2} , {#3} \succ
				}
			}
		}
	}
}
\newcommand{\locationDefinition}[1]{l = (rs,ps,#1)}

\usepackage{xspace}

\newcommand{\flowStar}{\texttt{Flow}$^*$\xspace}
\newcommand{\hypro}{\texttt{HyPro}\xspace}
\newcommand{\spaceex}{\texttt{SpaceEx}\xspace}

\newcommand{\keymaerax}{\texttt{Keymaera X}\xspace}
\newcommand{\phaver}{\texttt{PHAVer}\xspace}

\newcommand{\hycreate}{\texttt{HyCreate}}
\newcommand{\sapo}{\texttt{Sapo}}
\newcommand{\julia}{\texttt{JuliaReach}}

\def\gs{l}
\def\Type{{\it Type}}

\def\ds{{\it rs}}
\def\DS{{\it RS}}
\def\cs{{\it ps}}
\def\CS{{\it PS}}
\def\es{{\it es}}
\def\ES{{\it ES}}
\def\gs{l}
\def\gsDef{\gs=(\ds,\cs,\es)}

\def\NonurgentTransition{$\xRightarrow[]{}_N$}

\def\NonurgentTranSymbol{N}
\def\UrgentTranSymbol{U}

\def\NotUrgentTransition{\gs\nRightarrow_{\NonurgentTranSymbol}}

\def\UrgFlow{\textit{urg}'=1}
\def\UrgInv{\textit{urg} \le 0}
\def\ModeFlows{\textit{ModeFlows}}
\def\EventFlows{\textit{EventFlows}}
\def\ConstantFlows{\textit{FloatFlows}}
\def\ModeInvs{\textit{ModeInvs}}
\def\EventInvs{\textit{EventInvs}}

\def\EventResume{\textit{Resume}}
\def\EventEnqueue{\textit{Transfer}}
\def\EventEffectFunction{\textit{effect}}

\def\modes{\mathit{Modes}}

\def\Msg{\mathit{Msg}}

\def\ID{\mathit{ID}}

\def\Var{\mathit{Var}}
\def\Stmt{\mathit{Stmt}}

\def\Interval{\mathbb{IR}}

\def\map{\phi}
\def\timePassingTransition{\Rightarrow_{\it TP}}
\def\nonTimePassingTransition{\Rightarrow_{\it NTP}}
\def\messageTransition{\Rightarrow_m}
\def\statementTransition{\Rightarrow_s}
\def\tauTransition{\Rightarrow_\tau}

\def\body{\mathit{body}}
\def\int{\mathbb{N}}
\def\ET{\mathbb{ET}}

\let\emptyset\varnothing
\def\MessageServers{\it msgsrv}
\def\Modes{\it modes}

\def\HR{Hybrid Rebeca\space}

\def\m{\mathfrak{m}}

\def\bag{\mathfrak{b}}

\def\Valuation{\Sigma}
\def\valuation{\sigma}

\def\Dom{\mathit{Dom}}

\def\real{\mathbb{R}}

\def\TimeStep{\gamma}

\def\Jump{\mathcal{J}}

\newcommand{\HA}{\mathcal{H}}
\newcommand{\Loc}{\textit{Loc}}
\newcommand{\Init}{\textit{Init}}
\newcommand{\Inv}{\textit{Inv}}
\newcommand{\Flws}{\textit{Flow}}
\newcommand{\Jumps}{\textit{Jump}}
\newcommand{\guard}{\textit{guard}}
\newcommand{\reset}{\textit{reset}}
\newcommand{\source}{\textit{source}}
\newcommand{\target}{\textit{target}}

\newcommand{\R}{\mathbb{R}}

\journal{ Journal of Systems Architecture }
\begin{document}
\begin{frontmatter}

\title{Hybrid Rebeca Revisited}



\author[1]{Fatemeh Ghassemi}
\author[1]{Saeed Zhiany}
\author[1]{Nesa Abbasimoghadam}
\author[1]{Ali  Hodaei}
\author[1]{Ali Ataollahi}

\author[3]{J{\'o}zsef Kov{\'a}cs}
\author[3]{\newline Erika {\'A}brah{\'a}m}

\author[2]{Marjan Sirjani}

\affiliation[1]{organization={School of Electrical and Computer Engineering, University of Tehran},
city={Tehran},
country={Iran}
}

\affiliation[2]{organization={School of Innovation, Design and Engineering, Mälardalen University},
city={Västerås},
country={Sweden}
}

\affiliation[3]{organization={RWTH Aachen University},
city={Aachen},
country={Germany}
}




\begin{abstract}
  Hybrid Rebeca is a modeling framework for asynchronous event-based cyber-physical systems (CPSs). In this work, we extend Hybrid Rebeca to allow the modeling of non-deterministic time behavior. Besides the syntactical extension, we formalize the semantics of the extended language in terms of Timed Transition Systems, and adapt a reachability analysis algorithm originally designed for hybrid automata to be applicable to Hybrid Rebeca models. We prove the soundness of our approach and illustrate its applicability on a case study. 
  The case study demonstrates 
  that our dedicated algorithm is clearly superior to the alternative approach of transforming Hybrid Rebeca models to hybrid automata as an intermediate model and then applying the original reachability analysis method to this intermediate transformed models.

\end{abstract}


\end{frontmatter}


\section{Introduction}\label{sec::intro}
Cyber-Physical Systems (CPSs) are usually composed of distributed discrete controllers interacting with a physical environment via networked communication. Nowadays, CPSs are ubiquitous in safety-critical situations such as in automotive, manufacturing, and transportation domains. Thus it is important to analyse the safety of such \textit{hybrid systems} at design time before their implementation.

In order to enable a rigorous formal analysis,
CPSs are often modeled as \textit{hybrid automata}. Unfortunately, the problem of computing for a hybrid automaton all of its states that are reachable from some given initial states is undecidable \cite{alur1995algorithmic}.
Nevertheless, there are different algorithms to over-approximate the reachable state set within certain bounds (on the number of discrete steps, which is called the jump-bound, and the time horizon). If the over-approximated set does not contain any unsafe state, then we can conclude that the system is safe (within the given bounds). Most of these algorithms apply a technique called flowpipe construction \cite{Zha92}, as implemented in tools like \flowStar \cite{flow}, \hypro \cite{hyproToolPaper}, \spaceex \cite{spaceex}
, \hycreate \cite{HyCreate} 
, \sapo \cite{sapo}
and \julia \cite{julia}
.

However, modeling CPS as hybrid automata is not an easy task. To offer an alternative, in previous work we introduced Hybrid Rebeca \cite{jahandideh2018hybrid,SoSym} as an extension of (Timed) Rebeca \cite{sirjani2004modeling,aceto2011modelling}, in order to reliably  model asynchronously event-based CPSs. Rebeca (Reactive Object Language) provides an operational interpretation of the actor model through a Java-like syntax. Actors, called rebecs in Rebeca, are units of concurrency which can only communicate by asynchronous message passing. Each actor is equipped with a mailbox in which the received messages are buffered. The hybrid extension allows to model both the continuous and the discrete aspects of CPSs in a unified way by introducing physical rebecs as an extension of software rebecs with real-valued variables and modes in which the physical dynamics are specified. This way, Hybrid Rebeca provides a suitable level of abstraction for both cyber-physical components and the network. 


To verify the safety properties of a Hybrid Rebeca model, the state-of-the-art approach\cite{SoSym} is to transform the model into  a hybrid automaton and then inspect the reachability of an unsafe state in the hybrid automaton. 
In this approach, the reachability analysis of a Hybrid Rebeca model depends on the derivation of its corresponding hybrid automaton. The number of locations in the derived automata increases exponentially as the number of discrete variables in the software rebecs or the size of the rebec mailbox increases. 
Even though the time- and jump-bounded reachability analysis of the resulting hybrid automaton might not depend on all locations, in general the exponential model size makes it infeasible to verify complex systems this way.


\medskip

In this paper, we revise Hybrid Rebeca in both modeling and analysis aspects. For modeling a wide range of CPSs, supporting non-deterministic time behavior is required. For example, delays in communication or the completion of a cyber-computation have non-deterministic time behavior. Furthermore, physical rebecs may have non-deterministic behavior due to imprecision in measuring values that affect their behavior (this is modeled through non-deterministic behavior on staying in a mode or leaving it when a guard holds). For instance, we can model the behavior of a smart sensor that non-deterministically decreases the temperature when it exceeds a limit. 
To support the modeling of these aspects, we extend Hybrid Rebeca with modeling non-deterministic time behavior, realized by the notion of time intervals, and allowing non-deterministic behavior for physical rebecs. 
Besides the syntactical extension, we define the semantics of the extended Hybrid Rebeca language in terms of Timed Transition Systems (TTS). 

For the analysis, instead of generating hybrid automata and then computing their reachable states, we propose to compute the reachable states of Hybrid Rebeca models directly. Given a Hybrid Rebeca model and a set $I$ of initial states, we seek to over-approximate the set of the states that are reachable from $I$ in a bounded time horizon $[0,\Delta]$ and with at most $J$ jumps in the Hybrid Rebeca model, where $\Delta$ and $J$ are user-specified. 
To do so, we adapt the reachability analysis for hybrid automata \cite{flow} to iteratively compute successors for jumps as well as time-progressing transitions. As time advances, the real-valued variables of physical rebecs are approximated using an adaption of the flowpipe construction algorithm for hybrid automata. The termination of the algorithm is ensured by putting bounds on the time horizon as well as the jump depth.

In our new approach, which allows us to execute reachability analysis on the Hybrid Rebeca models directly, there is no need to generate large hybrid automata models as an intermediate representation. We demonstrate on a case study that the direct analysis Hybrid Rebeca models is more efficient than the alternative approach  via the transformation to hybrid automata.

Summarizing, our main contributions are:
\begin{itemize}
    \item extending Hybrid Rebeca with time non-deterministic behavior by allowing non-deterministic delays on computation, network communication, or physical mode management;
    \item formalizing the semantics of the extended Hybrid Rebeca models based on Timed Transition Systems; 
    \item designing an algorithm for the direct reachability analysis of Hybrid Rebeca models, adapting existing algorithms and tools for the reachability analysis of hybrid automata;
    \item proving the soundness of our direct reachability analysis algorithm;
    \item implementing the proposed algorithm and providing an experimental evaluation on a set of case studies, showing the efficiency of the new approach with the hybrid-automata-based technique on a case study.
\end{itemize}

\noindent \textit{Structure of the paper.}~
In Section \ref{sec::motiv} we motivate our work through an example. We recall preliminaries on hybrid automata and its flowpipe construction algorithm in Section \ref{sec::pre}. We extend the syntax and semantics of Hybrid Rebeca with non-deterministic timed behavior in Section \ref{sec::hyb}. We present our direct reachability algorithm and show its soundness in Section \ref{sec::hybImp}. We review related work in Section \ref{sec::related} and conclude the paper in Section \ref{sec::con}.

\begin{figure}[t]
	\centering
\begin{lstlisting}[language=HRebeca, multicols=3]
reactiveclass Controller(2){
  knownrebecs {Alarm alarm;}
  Controller(){}
  msgsrv control(float t){
    if (t<18.5)  alarm.notify()
    after (0.3,0.5);
  }
}

reactiveclass Alarm(2){
  statevars {int count;}
  Alarm(){}
  msgsrv notify() {
    count = 3;
    self.beep();
  }
  msgsrv beep() {    
    if (count>0){
        delay(0.2,0.4);
        self.beep();
        count = count-1;
    }
  }
}

physicalclass HeaterWithSensor(2){
  knownrebecs {Controller c;}
  statevars {float temp;}
  HeaterWithSensor(float temp_) {
    temp = temp_;
    setmode(off);
  }
  mode off {
    inv(temp>18){
      temp' = -1;
    }
    guard(temp<19){
      c.control(temp);
      setmode(on);
    }
  }
  mode on {
    inv(temp<23){
      temp' = 1;
    }
    guard(temp>22){
      setmode(off);
    }  
  }
}

main {
  Alarm a():();
  Controller c(a):();
  HeaterWithSensor hws(c):(20);
}
\end{lstlisting}
	\caption{A Hybrid Rebeca model of a room with a sensor-equipped heater, a controller, and an alarm mechanism.}
	\label{code::motiExample}
\end{figure}

\begin{figure}[t]
\centering
\scalebox{0.9}{\includegraphics[width=0.87\textwidth]{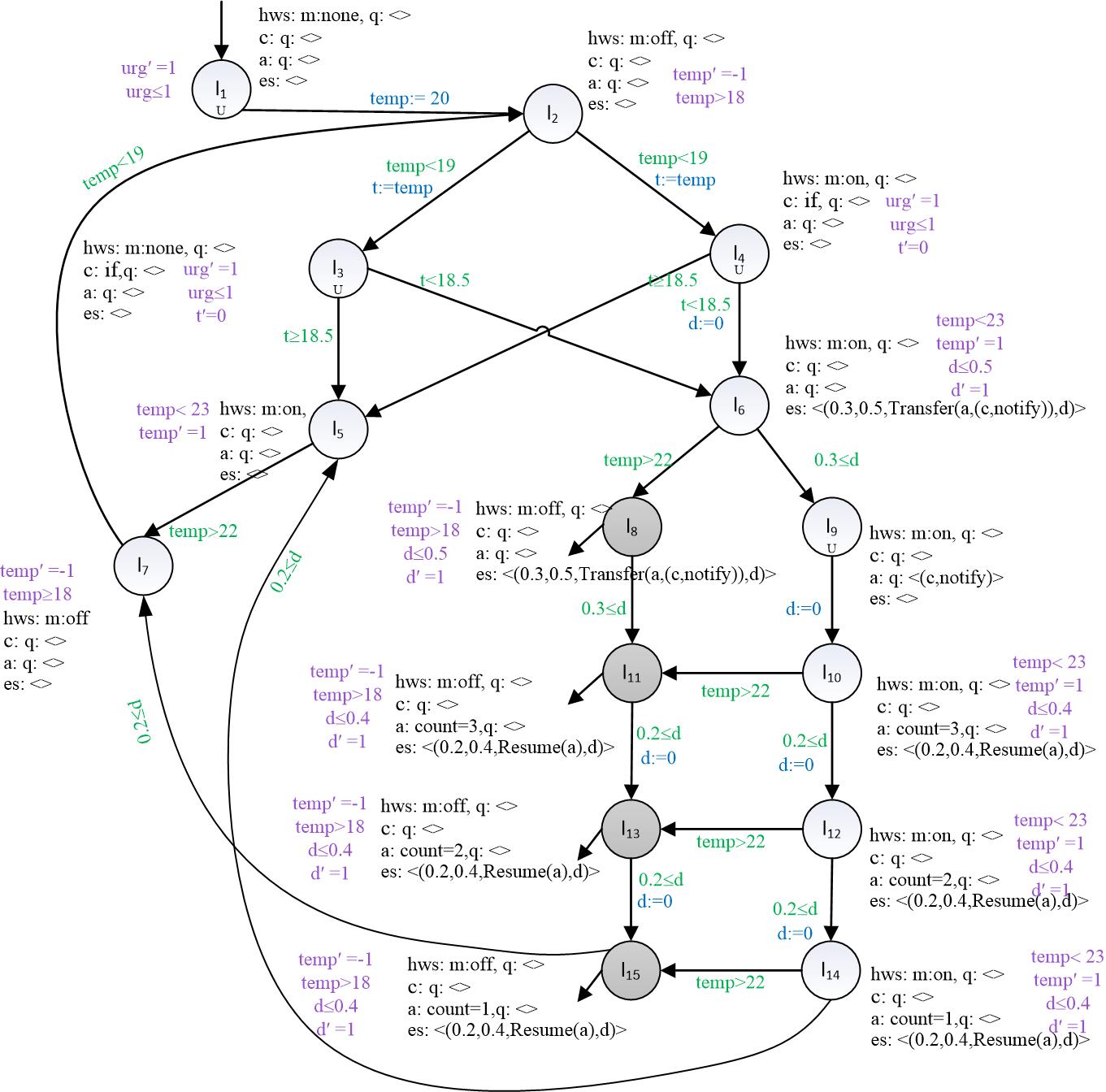}}
\caption{The monolithic hybrid automaton constructed for the Rebeca model given in Fig.~\ref{code::motiExample}; urgent locations, in which time does not pass, are indicated by `U'; invariants are shown in purple, guards in green,  resets in blue, and the mailbox and variable values of rebecs in black. \label{Fig::codeHA}}
\end{figure}

\section{Motivating Example}
\label{sec::motiv}

We start with a motivating example to illustrate the Hybrid Rebeca language, and the transformation process from Hybrid Rebeca to hybrid automata.

Consider a room equipped with a sensor-based heater and a controller. The heater has two modes $\emph{on}$ and $\emph{off}$; it switches from $\emph{on}$ to $\emph{off}$ at a non-deterministically chosen time point when the sensed temperature is between $22^\circ C$ and $23^\circ C$,  and from $\emph{off}$ to $\emph{on}$ between $18^\circ C$ and $19^\circ C$. Before switching to $\emph{on}$, it sends the temperature to the controller. If the received temperature is below $18.5^\circ C$, then the controller raises an alarm, triggering three beeps, with a non-deterministic delay from the interval $[0.2,0.4]$ before each beep. 
The communication from the heater to the controller is instantaneous, while communication from the controller to the alarm experiences a non-deterministic delay of $[0.3,0.5]$ seconds.

\smallskip

The Hybrid Rebeca model of the room is given in Fig. \ref{code::motiExample}. It consists of two software rebecs (\emph{Controller} and \emph{Alarm}) and a physical rebec (\emph{HeaterWithSensor}). The numeric arguments to the reactive class declarations denote the mailbox lengths. 

\smallskip

The corresponding monolithic hybrid automaton is shown in Figure \ref{Fig::codeHA}. The locations are defined by the local state of the actor instances (${\emph{hws}},c,a$) and the sequence of pending events ($es$), which are generated for delayed network or computation events. In the initial location $l_1$, which is urgent meaning that no time can pass in this location, the mailboxes of the three rebecs contain the statements of their constructors (not shown in the figure). Upon executing the statements of heater rebec's constructor, the mode of the heater is set to $\emph{off}$ and the temperature to $20$.
In location $l_2$, executing the mode statements (lines $33-41$) might trigger a $\emph{control}$ message being sent to the controller (line $38$). 
In that case, there is a non-deterministic race between the execution of the controller and the heater, leading to the locations $l_3$ and $l_4$. In location $l_4$, the controller processes the $\emph{control}$ message. As the concrete value of the transmitted temperature (modeled by $t$) is not defined, there is a branching between the cases when $t<18.5$ and $t \ge 18.5$,
leading to $l_5$ and $l_6$. Location $l_6$ indicates that $t<18.5$ and a message $\emph{notify}$ should be delivered to the alarm after a delay of $(0.3,0.5)$ (an event for this delay is added to the event list). 
Since the value of $\emph{temp}$ is not determined, there is again a non-deterministic behavior in location $l_6$: either the message $\emph{notify}$ is delivered leading to $l_9$ or the mode $\emph{on}$ triggers leading to $l_8$.

Given an analysis over the time-bound $[0,2]$, the location $l_8$ is not reachable.  Following a similar reasoning, the gray locations are all not reachable. 
There are also other locations accessible from locations $l_8$, $l_{11}$, $l_{13}$, and $l_{15}$, which are not shown because they are not reachable within the above time bound, but which nevertheless will be generated before the analysis starts.

\section{Preliminaries}\label{sec::pre}
Since we use a reachability analysis algorithm for hybrid automata to compute the reachable sets of a Hybrid Rebeca model, it is essential to first introduce hybrid automata and its respective reachability analysis algorithm.

Let $\int$ and $\real$ denote the natural (including 0) respectively real numbers.

\subsection{Hybrid Automata}\label{subsec::HA}
\emph{Hybrid automata (HA)}  \cite{alur1995algorithmic,henzinger2000theory} are a modeling formalism for systems whose evolution combines continuous dynamics with discrete steps, so called jumps.
Intuitively, a hybrid automaton can be viewed as an extension of a finite state machine with a finite set of real-valued variables. The values of these variables change continuously in a discrete state, called location, according to an ordinary differential equation (ODE). 

\begin{definition}\label{Def::HA}
	A \emph{hybrid automaton} is a tuple $\HA=(\Loc,\Var,
	\Jumps, \Flws, \Inv, \Init)$ with the following components:
	
	\begin{itemize}
		\item  $\Loc$ is a finite non-empty set of \emph{locations}.
		\item $\Var=\{x_1,\ldots,x_n\}$ is an ordered set of $n\in\int$ real-valued \emph{variables}, sometimes denoted in vector notation as $\vec{x}=(x_1,\ldots,x_n)$. 
		\item $\Jumps\subseteq \Loc\times
		2^{\real^n}\times (\real^n\rightarrow \real^n)\times \Loc$ is a finite set of \emph{jumps} (also called \emph{discrete transitions} or \emph{edges}) $e=(\ell,g,r,\ell')\in \Jumps$, with \emph{source location} $\source(e)=\ell$, \emph{target location} $\target(e)=\ell'$, 
		\emph{guard} $\guard(e)=g$, and \emph{reset} $\reset(e)=r$.               

		\item $\Flws$ is a function that assigns to each location a \emph{flow}, which is a set of ODEs of the form $\dot{\vec{x}}=f(\vec{x},t)$. 
		\item $\Inv: \Loc\rightarrow 2^{\real^n}$ is a labeling function that assigns to each location an \emph{invariant}. 
		\item $\Init\subset \Loc\times  \real^n$ is a set of \emph{initial states}. 
	\end{itemize}
\end{definition}

Assume in the following a HA $\HA=(\Loc,\Var,
\Jumps, \Flws, \Inv, \Init)$.
A \emph{state} of $\HA$ is a pair of $\langle \ell,\vec{x}\rangle$ where $\ell$ denotes the current location and $\vec{x}\in\real^n $ are the values of the variables.
An \emph{execution} of $\HA$ is a (finite or infinite) sequence $\langle \ell_0,\vec{x}_0\rangle,\langle \ell_1,\vec{x}_1\rangle,\ldots$ such that $\langle \ell_0,\vec{x}_0\rangle\in\Init$, $\vec{x}_0\in\Inv(\ell_0)$, and for each two successive states $\langle \ell,\vec{v}\rangle$ and $\langle \ell',\vec{v}'\rangle$ one of the following two cases holds:
\begin{itemize}
	\item Time progress: The location remains the same ($\ell=\ell'$), but the values of variables are updated with respect some time duration $t\ge 0$, such that $\vec{v}'=\varphi_f(\vec{v},t)$ with $\varphi_f$ being a solution to the flow in $\ell$ when starting in $\vec{v}$, and the invariant holds all the time, i.e. $\varphi_f(\vec{v},t')\in\Inv(\ell)$ for all $t'\in[0,t]$.
	\item Discrete jump: There exists a jump $(\ell,g,r,\ell')\in\Jumps$ whose guard is satisfied before the jump ($\vec{v}\in g$), and the successor valuation which results from the predecessor by applying the jump's reset ($\vec{v}'=r(\vec{v})$) satisfies the target location's invariant ($\vec{v}'\in \Inv(\ell')$).
\end{itemize}

\begin{figure}
	\centering
\begin{subfigure}[b]{0.48\textwidth}
		\includegraphics[width=0.9\textwidth]{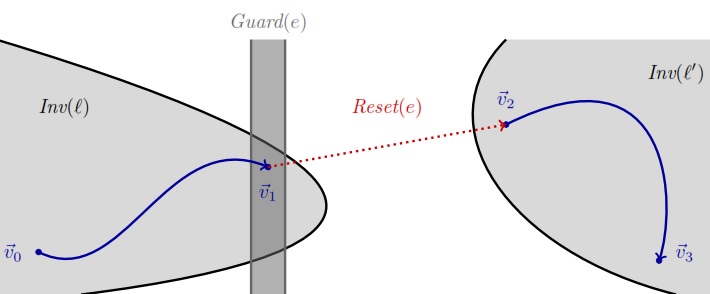}
\caption{an execution of a hybrid automaton }
\label{Fig::HAExbase}
\end{subfigure}
\hfill
\begin{subfigure}[b]{0.48\textwidth}
	\includegraphics[width=0.9\textwidth]{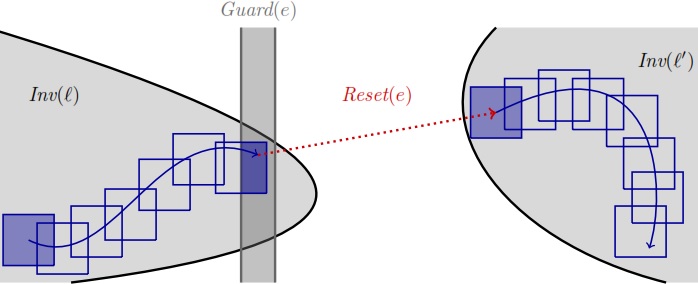}
    \caption{Flowpipe construction of the execution}
\label{Fig::HAExover}
\end{subfigure}
\caption{An example of an execution of a hybrid automaton and its flowpipe construction \cite{Chen}.}
\end{figure}

\noindent Figure \ref{Fig::HAExbase} illustrates an execution $\langle \ell,\vec{x}_0\rangle,\langle \ell,\vec{x}_1\rangle,\langle \ell',\vec{x}_2\rangle,\langle \ell',\vec{x}_3\rangle$, where $e\in\Jumps$ is a 
jump from $\ell$ to $\ell'$ \cite{Chen}.

\subsection{Hybrid Automata Reachability Analysis by Flowpipe Construction}

In general, the reachability problem for hybrid automata is undecidable. Therefore, instead of exact reachability, we aim at over-approximating the set of all reachable states using an approach called \emph{flowpipe construction}. As a data type to represent state sets in an over-approximative manner, the algorithm from \cite{Chen} uses Taylor models. The advantage of this approach is its support for non-linear ODEs. It has been implemented in \flowStar \cite{flow} and its code is available for reuse also in \hypro \cite{hyproToolPaper}. An algorithmic description of the method is shown in Algorithm \ref{alg::Chen}, where $\langle \ell,V\rangle$ denotes $\{\langle \ell,v\rangle\mid v\in V\}$ for $\ell\in\Loc$ and $V\subseteq \real^n$.
 
	\begin{algorithm}[t]
	
		\KwIn{a hybrid automaton $\HA$, a bounded time horizon $\Delta$, maximum jump $\mathcal{J}$}
		\KwOut{over-approximation of the reachable set of $\HA$ 
		}
		
		$\mathcal{R} \leftarrow \emptyset$;
		$Queue \leftarrow \emptyset$\;
		\ForEach{$i=1,\dots,m$}{
			$Queue.enqueue(\langle\langle \ell_i,V_i\rangle,0,\mathcal{J}\rangle)$ \tcc*{enqueue the initial sets}
		}
		\While{$Queue$ is not empty}{
			$\langle\langle \ell,V\rangle,t,j\rangle \leftarrow Queue.dequeue()$\;
			$F \leftarrow {\it ComputeFlowpipes}_l(V,\Delta - t)$ \tcc*{flowpipes in $l$ in time $\Delta-t$}
		  $\mathcal{R} \leftarrow \mathcal{R} \cup \{\langle \ell,\mathcal{F}\rangle\mid\mathcal{F} \in F\}$\;
		  \If{$j > 0$}{
			\ForEach{$(\ell,g,r,\ell') \in \Jumps$}{
				\ForEach{$\mathcal{F} \in F$}{
						$\mathcal{F_G} \leftarrow \mathcal{F}\cap g$ \tcc*{ flowpipe/guard intersection}
						\If{$\mathcal{F_G}\neq \emptyset$}{
							$\mathcal{F_R}\leftarrow r(\mathcal{F_G})$ \tcc*{ compute the reset mapping}
							\If{$\langle \ell',\mathcal{F_R}\rangle\not\subseteq \mathcal{R}$}{
								Compute the global starting time $t_R$ of $\mathcal{F_R}$\;
								$Queue.enqueue(\langle\langle \ell',\mathcal{F_R}\rangle,t_R,j-1\rangle)$\;
							}
						}
					}
				}
			}
		}
    		\Return $\mathcal{R}$\;
		\caption{Flowpipe construction for hybrid automata \cite{Chen}}	\label{alg::Chen}
	\end{algorithm}

Given a hybrid automaton $\HA$,  Algorithm \ref{alg::Chen} over-approximates the set of states reachable in a bounded time horizon $[0,\Delta]$ with at most $\mathcal{J}$ jumps, such that $\Delta$ and $\mathcal{J}$ are user specified.  The algorithm uses a queue to maintain the reachable sets computed up to a time point $t$ and the remaining number of jumps to be considered. The queue is initialized by the initial states of $\HA$, with the time progress initialized to $0$ (line $3$), and the remaining number of jumps to $\mathcal{J}$.  
For each state set in the queue, the algorithm over-approximates its successors, i.e. the set of all states reachable from it within the remaining time $[0,\Delta-t]$, by calling ${\it ComputeFlowpipes}_l$  (line $6$). This call returns a segmented flowpipe, i.e. a set of state sets called \emph{flowpipe segments}, whose union over-approximates the considered time-bounded reachability. Then, each (non-empty) flowpipe segment is intersected with the guard (line $11$) of each jump rooted in the current location, to inspect whether any jump is enabled in it. For enabled jumps, the reset is applied on the intersection (line $13$) to identify the successors, and the result is added to queue to be examined in the next iteration while the remaining time is updated and the remaining number of jumps is decreased. We illustrate the application of this algorithm on the execution of Figure \ref{Fig::HAExbase} and Figure \ref{Fig::HAExover} in which flowpipes are represented as boxes. 

\section{Hybrid Rebeca}
\label{sec::hyb}


\subsection{Syntax}

\noindent In Hybrid Rebeca, a hybrid system is modeled by two types of actors: 
\begin{itemize}
	\item \emph{Reactive classes} (or \emph{software classes}) are used to specify actors that have only discrete behavior. 
	\item   \emph{Physical classes} describe actors with a continuous behavior. Typically they are sensors and actuators in a system (e.g., a sensor that senses the temperature of the environment or the speed of a wheel in a car).
	
\end{itemize}

\def\HighlightColor{black}
\begin{figure}[t]
\fbox{\parbox{\columnwidth-5mm}{\vspace{-2mm}
			{\footnotesize
				\begin{align*}
				\mathrm{Model} &\Coloneqq~\langle\mathrm{RClass}\mid\textcolor{\HighlightColor}{\mathrm{PClass}}\rangle^+~\mathrm{Main}~\\
				\mathrm{Main} &\Coloneqq ~\mathsf{main}~ \{\mathrm{InstanceDcl}^*\} \\
				\mathrm{InstanceDcl} &\Coloneqq ~\mathrm{C}~ \mathrm{r}~(\mathrm{\langle  r\rangle}^* )~\colon(\mathrm{\langle c\rangle}^*)\\
				\mathrm{RClass} &\Coloneqq~ \textcolor{\HighlightColor}{\mathsf{reactiveclass}} ~\mathrm{C}~ \{\mathrm{KnownRebecs}~ \mathrm{Vars}~ \mathrm{MsgSrv}^* \}\\
				\mathrm{PClass} &\Coloneqq~ \textcolor{\HighlightColor}{\mathrm{physicalclass}} ~\mathrm{C}~ \{\mathrm{KnownRebecs}~ \mathrm{Vars}~ \mathrm{MsgSrv}^*~\textcolor{\HighlightColor}{\mathrm{Mode}^*} \}\\
				\mathrm{KnownRebecs} &\Coloneqq~ \mathsf{knownrebecs}~\{\mathrm{\langle Type~v\rangle}^* \}\\
				\mathrm{Vars} &\Coloneqq \mathsf{statevars}~
				\{\mathrm{\langle Type~v\rangle}^* \}\\
				\mathrm{MesgSrv} &\Coloneqq \mathsf{msgsrv} ~\mathrm{msg}~(\mathrm{\langle Type~v\rangle}^*)~\{\mathrm{Stmt}^* \}\\
				\mathrm{Mode} &\Coloneqq~ \textcolor{\HighlightColor}{\mathsf{mode}~ \mathrm{\mathfrak{m}}~\{\textcolor{\HighlightColor}{\mathsf{inv}(\mathrm{Expr})}~\{\langle\textcolor{\HighlightColor}{v'=\mathrm{Expr};}\rangle^+\}~\mathsf{guard} (\mathrm{Expr})~\{\mathrm{Stmt^*}\} \}}\\
				\mathrm{Stmt}
				&\Coloneqq~
				\mathrm{v}=\mathrm{Expr};\mid\mathsf{if}(\mathrm{Expr})~\mathrm{MSt}~[\mathsf{else}~\mathrm{MSt}]\mid \mathsf{delay}(c,c);\mid \mathsf{setmode}(\mathfrak{m});\mid\\
				&~~~~~~~\mathsf{r.msg(\langle \mathrm{Expr}\rangle ^*)~after(c,c); }\mid\textcolor{\HighlightColor}{\mathsf{r.SetMode(\mathrm{\mathfrak{m}})~after(c,c)}}\\
				\mathrm{MSt}
				&\Coloneqq~\{\mathrm{Stmt}^*\}\mid\mathrm{Stmt} \\
				\mathrm{Type} &\Coloneqq \mathsf{int}\mid \mathsf{float}\mid\mathsf{real}\\
				\mathrm{Expr}
				&\Coloneqq~c\mid v\mid\mathrm{Expr}~ \mathsf{op}~ \mathrm{Expr},~\mathsf{op}\in\{+,-,*,\wedge,\vee,<,\le,>,\ge,!=,==\} \mid\\
				&~~~~~~(\mathrm{Expr}) \mid\mathsf{!}(\mathrm{Expr)}
				\end{align*}
	}}}	
	\vspace{-2mm} \caption{Syntax of Hybrid Rebeca. Identifiers $C$, $msg$, $\m$, $v$, $c$, and $r$ denote the name of a class, message server, mode, variable, constant, and rebec, respectively.}
		\label{Fig::HybridRebecaGrammar}
\end{figure}

\noindent A reactive class' syntax contains a set of known rebecs, state variables, and message servers. Known rebecs are the rebecs that the instances of this class can send a message to. 
Message servers specify the behavior of a rebec upon handling messages. 
The main statements of Hybrid Rebeca's  language (used within the body of message servers) include variable assignment, setmode, conditional statement, delay, and sending messages. Each class definition must include a constructor message server with the same name as the class name. This message server is called upon instantiating the class to initialize the state variables and send initial messages. The \emph{self} keyword is used to send a message from a rebec to itself. This is helpful for defining iterative behaviors. 

A physical class is similar to a reactive class, except that it also includes the definition of physical modes. The structure of a mode is very similar to the locations in a hybrid automaton and includes the static condition (invariant), flow equations, the guard condition, and the corresponding statement block, called \textit{trigger}. Flow equations express ODEs of real variables. 
The conditional expressions of a static condition and a guard condition are defined by the reserved words \emph{inv} and \emph{guard}, respectively. The statement block of a guard condition specifies the behavior of a physical rebec in case of
satisfying the guard condition upon leaving the mode. The $\mathsf{setmode}$ statement specifies the next mode of the rebec. 
The mode \emph{none} is a special mode that is defined for all physical rebecs and shows the idle behavior of a rebec in which all the differential equations of the variables are zero. Activating this mode can be interpreted as stopping the physical behavior in physical rebecs. Other rebecs can change the mode of a physical rebec by sending the special message $\mathsf{SetMode}$ to it. This message server is defined for all the physical rebecs.

In \HR, there are three basic data types: int, float, and real. Int variables can only be used in reactive classes, real variables can only be used in physical classes, and float variables can be used in both. Mathematically, real and float values are the same; for each real variable, a flow differential equation is assigned to specify how its value changes over time. 


In Figure \ref{Fig::HybridRebecaGrammar}, we already extended the previous Hybrid Rebeca syntax to encorporate non-deterministic delays.
Firstly, we added the \emph{after} statement along with the message sending statements, which was removed in the previous version due to the separate definition of the network between the rebecs. In this statement as well as the \emph{delay} statement, it is possible to specify non-deterministic time behavior in the form of a continuous interval. 
As our focus in this work is on the reachability analysis of the extended \HR model, for simplicity we have removed the network entity; an extension with network entities is straightforward but technical (we had to maintain the local state of the network within the global states and subsume semantic rules to evolve the network states). In the main block, the rebecs are instantiated while the known rebecs and actual values of parameters of the constructors are passed via
two pairs of parenthesis. We remark that rebecs can be instantiated in any order irrespective of the dependency of a rebec to its known rebecs.


	

\subsection{Operational Semantics}
\label{subsec:operational-semantics}

Following the standard semantics of Timed Rebeca, we define the semantics of a Hybrid Rebeca model in terms of a \emph{Timed Transition System (TTS)}. The states in TTS are composed of the local states of reactive and physical rebecs and the global time interval. The global time interval denotes the times that rebecs can stay at the specified local states. 
Due to the presence of real variables that change over time, their values in the local states of physical rebecs and messages are defined by real intervals, similar to the global time. First, we introduce a few notations and auxiliary functions that are used in describing the formal semantics of Hybrid Rebeca.

\subsubsection{Notations}
Given a set $A$, we use $A^{*}$ to show a set of all finite sequences on elements of $A$.
For a sequence $a\in A^{*}$ with the length of $n$, the symbol $a_i$ denotes the element $i^{th}$ of the sequence, where $1 \leq i \leq n$. We use $x\in a$ if there exists $i$ such that $a_i=x$. 
An empty sequence is denoted by $\epsilon$ and $\langle h|T\rangle$ represents a sequence whose first element is $h \in A$ and $T \in {A}^*$ is the sequence of remaining elements. We use $a_1\frown a_2$ to denote a sequence achieved by appending $a_2\in A^{*}$ to $a_1\in A^{*}$, and $a\setminus x$ to remove an occurrence of $x$ from $a$ if it exists.

For a function $f:X \rightarrow Y$, we use the notation $f[\alpha \mapsto \beta]$ to denote the function $\{(a,b)\in f| a \neq \alpha\} \cup \{(\alpha, \beta)\}$, where $\alpha \mapsto \beta$ is an alternative notation for $(\alpha, \beta)$, $\alpha \in X$, and $\beta \in Y$. We use $\Dom(f)$ to denote the domain of the function $f$. 
%
A \emph{record type} is defined by ${({\it name}_1:T_1,\ldots,{\it name}_n:T_n)}$.
For each \emph{record} $\mathfrak{t} = (e_1,e_2,\dots,e_n)$, where $e_i\in T_i$, we use $\mathfrak{t}.{\it name}_i$ to denote the $i^{th}$-element of the tuple called ${\it name}_i$, i.e., $e_i$.
Furthermore, we use $\mathfrak{t}[e'/{\it name}_i]$ as a symbolic representation for the tuple achieved by replacing the element $\mathfrak{t}.{\it name}_i$ of $\mathfrak{t}$ by $e'$ in the record $\mathfrak{t}$.

An \emph{interval} is a set of all real numbers that are between two real endpoints $a,b\in\mathbb{R}$ such that $a\le b$.
We exploit the four types of conventional intervals, denoted as $(a,b)$, $(a,b]$, $[b,a)$, and $[a,b]$. 
The endpoints $a$ and $b$ are allowed to take the infinities $-\infty$ and $+\infty$ when the bound is strict. We call the intervals with non-strict bounds \textit{closed} intervals, and we denote the set of them by $\Interval$. For an interval $I =[a,b] \in \Interval$, we call $a$ the \textit{lower bound} and $b$ the \textit{upper bound} of $I$. We use $I_{\it low}$ and $I_{\it up}$ to denote the lower and upper bounds, respectively. A single value $a$ can be represented by the interval $[a,a]$.

For a set $\Var$ of variables ranging over values from $\textit{Val}$ (including boolean, $\int$, $\real$, and $\Interval$), a \emph{valuation} $\valuation: \Var \rightarrow \textit{Val}$ is a function that maps a value to each variable from its domain $\Dom(\valuation)=\Var$. The set of all valuations is denoted by $\Valuation$. For valuations $\valuation_1: \Var_1 \rightarrow \textit{Val}$ and $\valuation_2: \Var_2 \rightarrow \textit{Val}$ over disjoint domains $\Var_1\cap\Var_2=\emptyset$, we define their \emph{union} $\valuation_1\uplus \valuation_2$ to assign $\valuation_1(x)$ to each $x\in\Var_1$, and $\valuation_2(x)$ to each $x\in\Var_2$.

Let $\ID$ be a set of \emph{rebec identifiers}.  A \emph{message} is defined as a tuple $({\it sender}, m, p, {\it arrivalTime})$, where ${\it sender}\in \ID$ is the sender identifier, 
$m$ is the name of the message server, $p\in \Sigma$ is the parameter valuation defining the values of the parameters, and ${\it arrivalTime}\in \Interval$ is the time interval when the message is delivered to the rebec. We use $\Msg$ to denote the set of all messages.   Recall that we represent all messages that a physical rebec can send to a reactive rebec within the global time interval by one message by aggregating their arrival times into an interval and associating an interval to float parameters. We use $\body(x, m)$ to denote the body of the corresponding message server for the message name $m$ defined in the class type of the rebec with the identifier $x$, which is a sequence of statements. 
Each mode is defined by the tuple $(\mathfrak{m},i,f,g,a)$ where $\mathfrak{m}$ is the name of the mode, $i$, $f$, $g$  and $a$ are respectively invariant, flows, guard and trigger of the mode.

Each rebec has a mailbox in which it stores all the received messages. We represent the mailbox as a sequence of messages in the semantic states. Formally, a mailbox is defined as $\bag=\langle msg_1, msg_2, \dots , msg_n \rangle \in \Msg^*$ is a sequence of messages where $n \in \int$.

\subsubsection{Hybrid Rebeca Models and their Semantics}\label{subsec::Hmodel}\label{subsec::sos}
A Hybrid Rebeca model consists of the rebecs of the model. A software rebec consists of the definitions of its variables, message servers, and known rebecs. A physical rebec is defined like a software rebec plus the definitions of its modes. 

\begin{definition}[Hybrid Rebeca model]
A Hybrid Rebeca model is a tuple $(R_s,R_p)$ where $R_s$ and $R_p$ are disjoint sets of the software and physical rebecs in the model, respectively. The set $R = R_s \cup R_p$ denotes the set of all the rebecs in the model. A software rebec $r_s\in R_s$ and a physical rebec $r_p\in R_p$ with a unique identifier $i$, are defined by tuples $(i,V_{i},\MessageServers_{i},K_i)$ and $(i, V_{i},\MessageServers_{i},\Modes_{i},K_i)$, respectively, where $V_{i}$ is the set of its variables, $\MessageServers_{i}$ is the set of its message servers, $K_i\subseteq R$ is the set of its known rebecs, and $\Modes_{i}$ is the set of modes. 

\end{definition}

\noindent 
Rebecs respond to the expiration of a physical mode or taking a message from their message queues. A physical mode expires when its guard holds, then the trigger of the physical mode is executed. Upon taking a message, the rebec processes it by executing its corresponding message server. 
We define the following TTS-based operational semantics for Hybrid Rebeca models.

\begin{definition} [Timed Transition System for a Hybrid Rebeca model]
Given a Hybrid Rebeca model  $\mathcal{M}=(R_s,R_p)$, its formal semantics is defined by the TTS ${\it TTS}(\mathcal{M})=(S, \rightarrow, s_0)$, where the set of global states $S$, transitions $\rightarrow$, and the initial state $s_0$ are defined in the following.
\end{definition}

\subsubsection{Global States }
\label{subsec::FormalSemantics}


The global states of a Hybrid Rebeca model are defined by the local states of its rebecs and the global time, denoted by the record $(RS:rs,PS:ps,GT:gt)$. The elements $rs$ and $ps$ define the local states of reactive and physical rebecs, respectively, by mapping a given rebec identifier $x \in \ID$ to its local state, and $gt$ is the global time. 

\begin{definition}[Local state of a reactive rebec]\label{DefReactiveState}
The local state of a reactive rebec is defined as a tuple $(\valuation,\bag,st,r) \in \Valuation \times \Msg^* \times \Stmt^* \times \Interval$, where $\valuation \in \Valuation$ is the valuation function that maps a given $v \in \Var$ to its value, $\bag$ is the rebec's mailbox, $st \in \Stmt^*$ is the sequence of statements of current message server the rebec must execute them, and $r$ donates the rebec's resume time.
\end{definition}

\begin{definition}[State of a physical rebec]\label{DefPhysicalState}
The state of a physical rebec is defined as a tuple $(\sigma,\bag,st,\m) \in \Valuation \times \Msg^* \times \Stmt^*  \times \modes$  where $\sigma \in \Valuation$ is the valuation function that maps a given $v \in \Var$ to its interval, $\bag$ is the rebec's mailbox, $st \in \Stmt^*$ is the sequence of statements of current message server the rebec must execute them, and  $\m$ donates the rebec's current active mode.
\end{definition}


\subsubsection{Transitions}
There are two main types of transitions in the \HR semantics: Time Passing ($\it TP$) and Non-Time Passing ($\it NTP$) transitions. The former causes global time progress, denoted by $\timePassingTransition$, while the latter does not, denoted by $\nonTimePassingTransition$. Non-time progressing transitions show taking a message by a rebec from its mailbox ($\messageTransition$), executing a statement by a rebec ($\statementTransition$), or resuming a rebec ($\tauTransition$). 
%

\subsubsection*{Take message rule}
Upon taking a message by a rebec from its mailbox, the statements of its corresponding message server should be executed. A reactive rebec can take a message from its mailbox, when (1) it has no statement to execute from its previously taken message, (2) it is not suspended (in this case the resume time is denoted by $\bot$
), and (3) there exists a message in the mailbox that the lower bound of the global time is greater than the lower bound of the message's arrival time. If the upper bound of the global time is less than the upper bound of the message's arrival time, there is a non-deterministic behavior on either handling the message assuming that the message has been received in a time within the global time interval, or assuming that the message will be arrived in future. Thus, for those arrival times within the global time interval, the message is handled and removed from the mailbox. However, for the arrival time out of the global time interval, the message remains in the mailbox.
The rule for these two scenarios is given below: 

\begin{equation*}
\label{Sos::ReactiveTakeMessage}
\Rule
{
	\begin{array}{c}
	\locationDefinition{[t_1,t_2)} \land l.RS(x)=(\sigma,\bag,\epsilon,\bot)\,\land \\ msg=(s,m,p,[t_3,t_4) ) \in \bag \land t_3 \le t_1
	\end{array}
}
{
	l \messageTransition l\left[rs\left[
	\begin{array}{clc}
	x\mapsto(
	&\sigma \uplus p,
	\bag \setminus msg,body(x,m),\bot&)
	\end{array}
	\right]/RS\right]
}
\end{equation*}

\begin{equation*}
\label{Sos::ReactiveNotTakeMessage}
\Rule
{
	\begin{array}{c}
	\locationDefinition{[t_1,t_2)} \land l.RS(x)=(\sigma,\bag,\epsilon,\bot)~\land \\ msg=(s,r,m,p,[t_3,t_4)) \in \bag \, \land\, t_3 \le t_1  
    \land\, t_2 < t_4
	\end{array}
}
{
	l \messageTransition l\left[rs\left[
	\begin{array}{clc}
	x\mapsto(
	&\sigma,(\bag \setminus msg)\frown (s,r,m,p,[t_2,t_4)),\epsilon,\bot)&)
	\end{array}
	\right]/RS\right]
}
\end{equation*}

\noindent  

A physical rebec can take a message if the above-mentioned conditions hold and additionally the message is not $\it SetMode$ and the mode of rebec is not $\it none$. 
If the message name is $\it SetMode$, no statement is added for execution while the active mode of rebec is updated to the one indicated by the message.

%

\subsubsection*{Resume and Postpone rules} A reactive rebec is suspended when it executes a delay statement. When the lower bounds of global and resume time intervals are equal, the rebec can resume its execution. 

\begin{equation*}
\label{Sos::ResumeComplementary}
\Rule
{
	\begin{array}{c}
	\locationDefinition{[t_1,t_2)} \land l.RS(x)=(\sigma,\bag,st,[t_3,t_4))~\land 	t_1 = t_3 
	\end{array}
}
{
	l \tauTransition l[rs[x\mapsto(\sigma,\bag,st,\bot)]/RS]
}
\end{equation*}

\noindent When a rebec can be resumed while the upper bound of the global time interval is less than the upper bound of a rebec's resume time, then a rebec has a non-deterministic behavior: either the rebec resumes its statement execution (above rule) or postpone resuming the rebec to the future as explained by the  SOS rule in below. 
\begin{equation*}
\label{Sos::ResumeComplementary}
\Rule
{
	\begin{array}{c}
	\locationDefinition{[t_1,t_2)} \land l.RS(x)=(\sigma,\bag,st,[t_3,t_4))~\land 	t_1 = t_3 \land t_2 <  t_4
	\end{array}
}
{
	l \tauTransition l[rs[x\mapsto(\sigma,\bag,st,[t_2,t_4))]/RS]
}
\end{equation*}


\subsubsection*{Delay Statement} The execution of a reactive rebec is suspended upon executing a delay statement. This statement affects the resume time of the rebec.
\begin{equation*}
\label{Sos::Delay}
\Rule
{
	\begin{array}{c}
	\locationDefinition{[t_1,t_2)} \land l.RS(x)=(\sigma,\bag,\langle delay(d_1,d_2) \mid st\rangle,\bot)
	\end{array}
}
{
	l \Rightarrow_s l[rs[x\mapsto(\sigma,\bag,st,[t_1 + d_1,t_2 + d_2))]/RS]
}
\end{equation*}

\subsubsection*{Send Statement}
The SOS rule for the send statement expresses that by executing the send statement, a message is inserted into the mailbox of the receiver rebec while its arrival time is defined based on the $\mathsf{after}$ construct. The rules for the send statement when the sending or receiving rebec is a physical rebec are the same.
\begin{equation*}
\label{Sos::RtoRSendMessage}
\Rule
{
	\begin{array}{c}
	\locationDefinition{[t_1,t_2)} \land l.RS(x)=(\sigma,\bag,\langle y.m(p)~\it{after}(d_1,d_2) \mid st\rangle,\bot)
    ~\land \\
	l.RS(y)=(\sigma',\bag',st',rt) 
	\end{array}
}
{
	l \Rightarrow_s l\left[
	\begin{array}{rcr}
	rs[&
	x\mapsto(\sigma,\bag,st,\bot)&]/RS\\
	rs[&y\mapsto (\sigma',\bag'\frown(x,y,m,p,[t_1+d_1,t_2+d_2)),st',rt)&]/RS
	\end{array}
	\right]
}
\end{equation*}

\subsubsection*{Assignment}
The SOS rule for the assignment statement specifies how the internal state of an actor is modified upon executing an assignment. The auxiliary function $\it eval$ computes the value of an expression based on the given variable valuation. 
\begin{equation*}
\label{Sos::Assignment}
\Rule
{
\begin{array}{c}
\locationDefinition{[t_1,t_2)} \land l.RS(x) = (\sigma, \bag, \langle v = expr \mid st \rangle,\bot) 
\end{array}
}
{
l \Rightarrow_s l\left[
\begin{array}{rcr}
rs[&
x \mapsto (\sigma[\textit{v} \mapsto {\it eval}(expr,\sigma)], \bag, st, \bot )&]/RS
\end{array}
\right]
}
\end{equation*}

\subsubsection*{Conditional Statement}
The SOS rule for the conditional statement specifies how an actor's state evolves based on evaluating a boolean condition which can be in the form of ${\it exp}r<c$, ${\it expr}>=c$, and their conjunctions and disjunctions. As the value of an expression may be an interval, there maybe values within the interval that satisfy the condition or there maybe values that dissatisfy the condition. In this case, the conditional statement generates two transitions for executing the ``then'' and ``else'' parts. If the value of the expression only satisfies or dissatisfies the condition only one transition for executing the ``then'' or ``else'' part is derived. The following rules consider the condition expression as ${\it expr}<c$ while the rules with more complex conditions can be defined similarly. 

\[
\Rule
{
    \begin{array}{c}
     l = (rs, ps, [ t_1, t_2 )) \land l.RS(x) = (\sigma, b, \langle \text{if}~(expr<c)~\text{then}~st_1~\text{else}~st_2 \mid st \rangle, \bot)~\land\\
      {\it eval}(expr,\sigma)_{\it up} <  c
    \end{array}
}
{
    l \Rightarrow_s l\left[
    \begin{array}{rcr}
     rs[&
     x \mapsto (\sigma, b, st_1\frown st, \bot)&]/RS
     \end{array}
     \right] 
}
\]

\[
\Rule
{
    \begin{array}{c}
    l = (rs, ps, [ t_1, t_2 )) \land l.RS(x) = (\sigma, b, \langle \text{if}~(expr<c)~\text{then}~st_1~\text{else}~st_2 \mid st \rangle, \bot)~\land\\
     {\it eval}(expr,\sigma)_{\it up} \ge  c
    \end{array}
}
{
    l \Rightarrow_s l\left[
    \begin{array}{rcr}
    rs[&
    x \mapsto (\sigma, b, st_2\frown st, \bot)&]/RS
    \end{array}
    \right] 
}
\]
For the expression whose values either satisfy or dissatisfy the condition, we have two possible transitions:

\[
\Rule
{
    \begin{array}{c}
    l = (rs, ps, [ t_1, t_2 )) \land l.RS(x) = (\sigma, b, \langle \text{if}~(expr<c)~\text{then}~st_1~\text{else}~st_2 \mid st \rangle,\bot)
    ~\land \\ {\it eval}(expr,\sigma)_{\it low}  \le  c < {\it eval}(expr,\sigma)_{\it up} 
    \end{array}
}
{ \begin{array}{c}
    l \Rightarrow_s l\left[
    \begin{array}{rcr}
    rs[&
    x \mapsto (\sigma, b, st_1\frown st, \bot)&]/RS
    \end{array}
    \right] \\
    l \Rightarrow_s l\left[
    \begin{array}{rcr}
    rs[&
    x \mapsto (\sigma, b, st_2\frown st, \bot))&]/RS
    \end{array}
    \right]
    \end{array}
}
\]

\subsubsection*{Time progress rule} The global time is advanced when no rebec can take a message or execute a statement. To explain when and how much the global time interval progresses, we first explain the \emph{earliest event time} concept.  The global time progresses according to the earliest event time by which a change is possible either in the local state of a reactive rebec or in the flow of a physical rebec. 
A change, so called an \emph{event}, can happen when a rebec can take a message from its mailbox, or the guard condition of a physical rebec in a mode is enabled, or a reactive rebec can resume its executions. Based on these three events, we consider the following \emph{time event} set, denoted by $\ET$, containing:
\begin{itemize}
	\item The lower and upper bounds of the arrival times of  messages in the mailbox of rebecs that can be taken.
	\item The lower and upper bounds of the resume times of suspended reactive rebecs.
	\item The lower and upper 
    bounds of the time interval that a physical rebec's guard is enabled to be taken. We assume that there is a function ${\it trigInterval}: \ID \times \Valuation \times \modes \rightarrow \Interval$ that can calculate the time interval that a physical rebec can be triggered in its mode $\m \in Modes$ 
	according to its current valuation function $\valuation \in \Valuation$
\end{itemize}

We define a total order relation on the elements of $\ET$ irrespective to their closeness and openness. We define the $i^{\it th}$ earliest event time ${\it EET}_i$ as the $i^{\it th}$ smallest value in this set. The following rule makes progress to the global time interval computed by the auxiliary function ${\it progressTime}$ with respect to ${\it EET}_1$ and ${\it EET}_2$. 

\[ 
\label{Sos::TimeProgress}
\Rule
{
	\begin{array}{c}
	l \not\nonTimePassingTransition \land~ \locationDefinition{[t_1,t_2)} \land 
	gt'={\it progressTime}(t_1,t_2,{\it EET}_1,{\it EET}_2)~\land\\
    ps'\in{\it updatePhysicalRebecs}(ps,gt')
	\end{array}
}
{
	l \timePassingTransition l[
	ps'/PS, gt' /GT]
}	
\]
Upon advancement of global time, the real variables of all physical rebecs are updated in case the invariant of their mode holds. Due to the non-determinsim behavior of physical rebecs in staying in a mode or leaving the mode, there are several ways that physical rebecs can be updated. The auxiliary function ${\it updatePhysicalRebecs}(ps,gt')$ computes all possible updates of physical rebecs. 
For physical rebecs like $x$ such that $ps(x)=(\sigma,\bag, st,\mathfrak{m})$, whose trigger intervals are a subset of $gt'$, i.e., ${\it trigInterval}(x,\sigma,\mathfrak{m})\subseteq gt'$, the remaining statements of these physical rebecs can be also updated to their guard's trigger. For those intervals where taking a message and triggering a mode are both possible in a physical rebec, we give priority to the trigger event. The auxiliary function ${\it progressTime}: \mathbb{R} \times \mathbb{R} \times \mathbb{R} \times \mathbb{R} \rightarrow \text{Interval}$ finds the new global time interval according to the given values:

\[
{\it progressTime}(t_1,t_2,t_3,t_4) = 
\begin{cases}
    [t_3, t_2+(t_3-t_1)), & \text{if } t_3 < t_2 \\
    [t_2, t_3), & t_3>t_2 \\
    [t_3,t_4), & t_2=t_3 
\end{cases}
\]


\subsubsection{Initial State}
The initial state is achieved by initializing the state variables of rebecs to their default values, i. e., zero, and then executing the statements of constructors of rebecs in the order instantiated in the main block.

\begin{example}
\label{example::ReachabilityAnalysisUsingSOS}
The derived transition system of our running example is shown in Figure ~\ref{Fig::codeTTS}. The initial state is $s_0$ in which the rebecs have executed their constructors in the order that they have instantiated in the main block. All rebecs have an empty mailbox and the physical rebec $\emph{hws}$ is in the mode $\emph{off}$ and the variable $\emph{temp}$ is initialized to $[20,20]$. As the controller and alarm have no message in their mailboxes, only the time-progress rule can be applied. The guard of mode $\emph{off}$ holds when the global time is $(1,2)$ and the temperature is $(18,19)$. By the time-progress rule, the global time is advanced to $[0,1)$. 
%
%
We call the resulting state $s_1$ in which no event can occur as the guard condition of the mode $\emph{off}$ is not enabled. In this state, only the time-progress rule can be applied again leading to the advancement of the global time to $[1,2)$. When the global time is $[1,2)$, the temperature is $(18,19]$ and both the invariant and guard of this mode are satisfied. Either the sensor can stay in this mode leading to the state $s_3$ or leave the mode by executing its trigger. Upon executing the trigger, a $\emph{control}$ message is immediately sent to the controller and the sensor switches to the mode $\emph{on}$. Upon processing the $\emph{control}$, there is a non-deterministic behavior as there are values for temperatures that both dissatisfy and satisfy the if-condition resulting in the states $s_5$ and $s_6$, respectively. When the condition is satisfied, a $\emph{notify}$ message is sent to the alarm. Due to the non-deterministic delay on the communication between the controller and alarm, the arrival time of the $\emph{notify}$ message is set to  $[1.3,2.5)$. 
In the state $s_6$, the message can not be taken and the time-progress rule is applied which advances the lower-bound of the global time to $1.3$, i.e., the lower bound of the message arrival time. Then,  there is a non-deterministic behavior in the state $s_6$. Either the message has arrived between $[1.3,2.3)$ and will be process by the alarm or it will be arrived at $[2.3,2.5)$. Upon taking the message by the alarm, it sets  $\emph{count}$ to $3$ and sends a message $\emph{beep}$ to itself. It then take this message and waits for $[0.2,0,4]$, leading to the state $s_7$ with the resume time set to $[1.5,2.7)$. Postponing the message arrival to $[2.3,2.5)$ results in the state $s_8$. In the state $s_7$, the time-progress rule can only be applied and the global time is advanced to $[1.5, 2.5)$. There is a non-deterministic behavior on suspending the alarm; either the alarm will be resumed leading to the state $s_{10}$ or its suspension is postponed to $[2.5,2.7)$ in the state $s_{11}$. Upon resuming the alarm in the state $s_{10}$, it sends another $\emph{beep}$ message for the second time to itself and sets $\emph{count}$ to $2$. Alaram takes this message and waits for a delay of $(0.2,0.4)$. The states $s_7$, $s_{10}$ and $s_{12}$ have a similar behavior. After handling three $\emph{beep}$ messages in $s_{14}$, the only rule that can be applied is the time-progress rule. The time is advanced to $[2.9,3)$ by the time bound of $3$.  

\begin{figure}[t]
\centering
\includegraphics[width=0.85\textwidth]{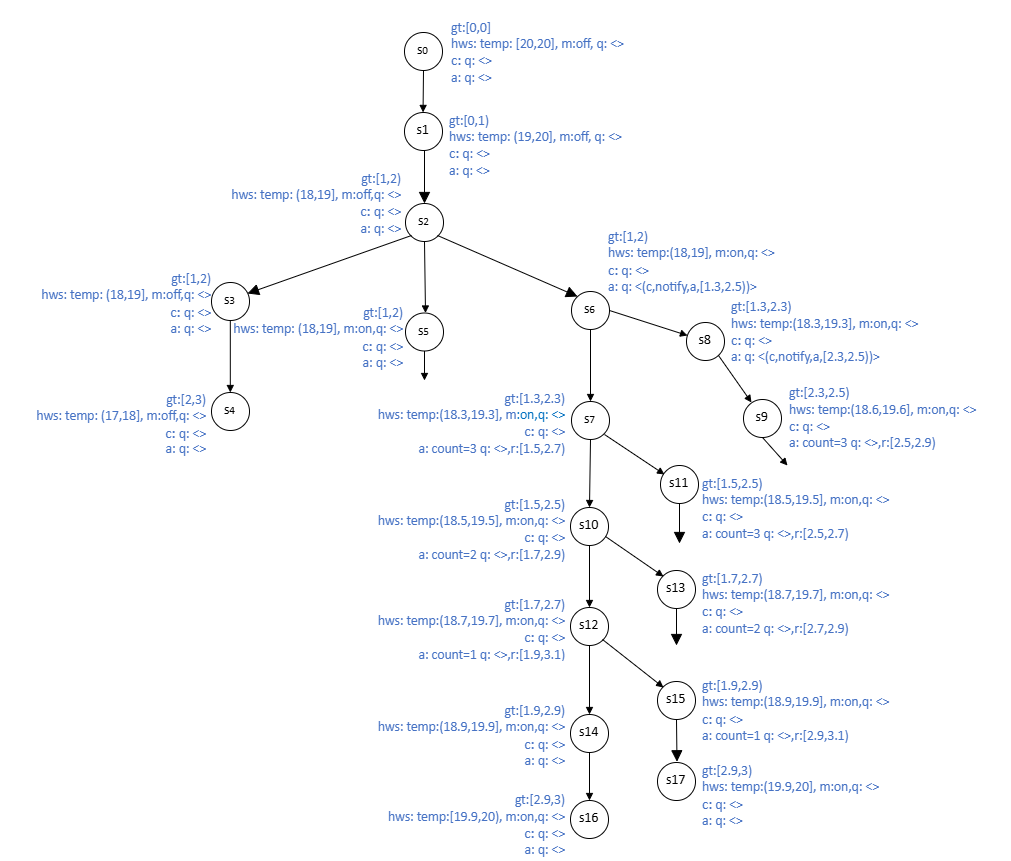}
\caption{The corresponding timed transition system of the Rebeca model given in Fig. \ref{code::motiExample} \label{Fig::codeTTS}}
\end{figure}	
\end{example}

\section{Reachability Analysis of Rebeca Models}\label{sec::hybImp}

We introduced the semantic model of \HR in Section \ref{sec::hyb}, which relies on the function $\it TrigInterval$ in the SOS rule of time progress to solve differential equations. As solving all non-linear ODEs is not decidable, 
we employ numerical methods to replace such a function by proposing an algorithm for calculating the over-approximation of the reachable states of a given \HR model within a time-bounded interval $[0,\Delta]$ and maximal depth $\Jump$, inspired by Algorithm \ref{alg::Chen}. 

{\small
\begin{algorithm}[!h]
\caption{ over-approximation reachability analysis}
\KwIn{a \HR model $\mathcal{HR}$, a bounded time horizon $\Delta$, maximum jump depth $\mathcal{J}$, and step size $\TimeStep$
}
\KwOut{over-approximation of the reachable set of $\mathcal{HR}$'s states
}
\label{alg::HybridRebecaReachabilityAnalysisAlgorithm}

$Queue \leftarrow \emptyset$\;
$s_0\leftarrow{\it makeInitialState}(\mathcal{HR})$\; 
$R \leftarrow {\it executeNTPSOS}(s_0)$ \;
\ForEach{$r \in R$} {
	$Queue.equeue(r,\mathcal{J})$
}
\While{$Queue$ is not empty} { 
$\{s:(RS,PS,GT),\mathcal{J}\} \leftarrow Queue.dequeue()$\;
\If{$(\mathcal{J}==0)\vee (GT_{\it low}==\Delta)$}{continue;}

$GT'\leftarrow {\it progressTime}(GT_{\it low},GT_{\it up},{\it EET}_1,{\it EET}_2)$ \tcp*{ where $\ET={\it events}(s)\cup \{GT_{up}+\gamma,\Delta\}$}
$ {\it ComputeFlowpipes}_{{\it flows}(PS)}({\it vals}(PS),\Delta- GT'_{low})$\;
  
$PS'\leftarrow {\it updatePhysicalRebecs}(PS,GT')$\;

${\it intersect}({\it inv}(PS'),{\it vals}(PS'),GT')$\;
\If{${\it vals}(PS')$ is empty}
  {continue;}
$V\leftarrow\{(RS,PS',GT')\}$\;

\ForEach{$p\in \Dom(PS')$} {
	\If{${\it guard}(p,PS'(p).\mathfrak{m})$ is satisfiable\tcp*{ $PS'(p).\mathfrak{m}$ denotes the mode of rebec $p\in\ID$}} 
	{\ForEach{$v\in V$}
		{$v' \leftarrow {\it new~GlobalState}(v)$\;
		$v'.PS'(p).st \leftarrow {\it Trigger}(p,ps.\mathfrak{m})$\tcp*{${\it Trigger}(p,ps.\mathfrak{m})$ denotes the mode statements}
		$v'.PS'(p).\mathfrak{m} \leftarrow {\it none}$\;
		$V.{\it add}(v')$\;
		 }
	}
}
	\ForEach{$v\in V$}
	{
		$S = {\it executeNTPSOS}(v)$\;
		$R \leftarrow R \cup S$\;
		\ForEach{$s\in S$}
		{
			$Queue.equeue(s,\mathcal{J}-1)$;
		}
	}
 
}
\Return $R$;
\end{algorithm}
}

Similar to Algorithm \ref{alg::Chen}, we use a queue to maintain the reachable sets with the number of jumps. The initial state $s_0$ is constructed by calling $\it makeInitialState$ which initializes the local variables of rebecs to their default values and executes the constructor of the rebecs in the order that they are instantiated in the main block. We find the set of reachable states as the result of non-time passing transitions by calling the function ${\it executeNTPSOS}(s_0)$. This function iteratively applies the SOS rules for non-time passing transitions given in Section \ref{subsec:operational-semantics} to $s_0$ and then the new reachable states until no such SOS rules can be applied further. The resulting reachable states from $s_0$ initialize the set of reachable states $R$ (line $3$). Then, we insert the items of $R$ (paired with $\mathcal{J}$) into $\it Queue$ to find the reachable state from them. In the while-loop, we dequeue an item and find the reachable states from it if $(\mathcal{J}> 0)\wedge (GT_{\it low}\neq \Delta)$. We advance the global time interval by calling ${\it progressTime}$ (line $10$) which computes the next global time based on the current global time, events of the state $s$, denoted by ${\it events}(s)$, and $GT_{\it up}+\gamma$, the time bound $\Delta$. This function was also used in the time progress SOS rule in Section \ref{subsec::sos}. By ${\it events}(s)$, we mean the lower/upper bounds of arrival time of messages in the mailbox of rebecs and the lower/upper bounds of the resume time of rebecs. To over-approximate the value of real variables of physical rebecs for the remaining time bound $\Delta-GT'_{\it low}$, we call the \hypro code, i.e., ${\it ComputeFlowpipes}_{{\it flows}(PS)}({\it vals}(PS),\gamma,\Delta- GT'_{low})$ where ${\it flows}(PS)$ collects the ODEs of the physical rebecs.  We store these values for each stepsize $\gamma$ within the bound for the current modes of physical rebecs ${\it flows}(PS)$ and the current values of real variables, denoted by ${\it vals}(PS)$.

We update the values of physical rebecs's variables based on the next global time by calling $\it updatePhysicalRebecs$. This function integrates the stored values of those stepsizes that intersect with the global time interval.  We intersect the invariants of all physical rebecs, denoted by ${\it inv}(PS')$ 
with the values of the variables updating $PS'$ and $GT'$ based on this intersection. We find those physical rebecs that can do a discrete jump (line $18$). We initialize the set $V$ to $\{(PS',RS,GT')\}$, the new state reached upon advancing the global time. For each physical rebec that the guard of its mode is satisfiable by some value, we make a copy of the states of $V$ in which the guard of 
the rebec triggers by updating its statements to the trigger statements (lines $20-23$). For each state $v\in V$, we find the reachable states from $v$ by calling the function ${\it executeNTPSOS}(v)$, update the reachable states $R$, and inserting the new reachable state for reachability analysis into the $\it Queue$ (lines $24-28$).

\begin{figure}
	\centering
	\includegraphics[width=\textwidth]{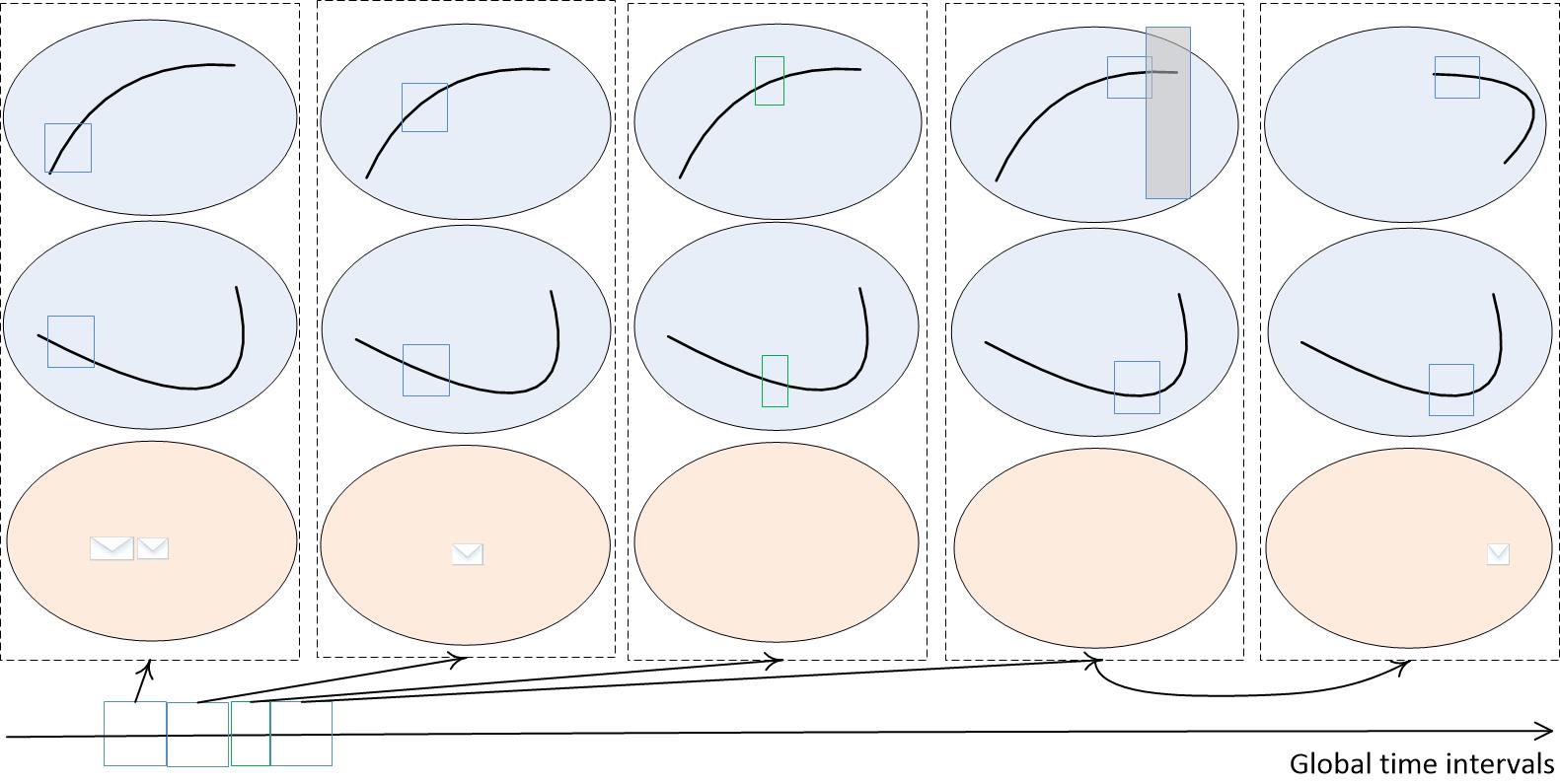}
	\caption{The intuition of reachability algorithm for a model of two physical rebecs and one reactive rebec.}
	\label{Fig::RebeEx}
\end{figure}

Figure \ref{Fig::RebeEx} provides some intuition for our algorithm. The intervals of global times are not all equal to $\gamma$ and the corresponding global state to each global time is shown. The global states consist of two physical rebecs and one reactive rebec. In the first two iterations of the while-loop, the intervals for global time are defined by $\gamma$. In the first iteration only the local states of physical rebec are updated (in line $13$), while in the second iteration the reactive rebec can handle one of its messages. The upper bound of the arrival time of the only message in the mailbox of the reactive rebec defines the upper bound of global time for the third iteration (the green box). By taking the message, the local states of all rebecs are updated (line $25$). In the fourth iteration, there is a possible change of mode by one of the rebecs which will send a message to the reactive rebec (the mode change of the rebec is identified by line $18$ in the algorithm, and its send message statement and change mode are due to the lines $20-23$). Intuitively, a state is inserted into $\it Queue$ when it has no enabled non-time passing transition. After the time progresses, the local states of physical and software rebecs are updated accordingly, by applying all enabled non-time passing transitions again.



\begin{example} \label{exm::hybridHeaterCaseStudy}
To illustrate the applicability of our reachability algorithm, we apply it to our motivating example. The resulting state space is shown in Figure \ref{Fig::reachalgo}. 

We start the analysis from the initial state $s_0$ in which the rebecs have executed their constructors. We assume that time bound $\Delta$, maximum jump $\mathcal{J}$, and $\gamma$ are set to $3$, $10$, and $0.5$, respectively. By calling ${\emph{executeNTPSOS}}$ (line $3$), no new state is generated and only $s_0$ is inserted into $\emph{ Queue}$. In the first and second iterations, the global time is advanced for $0.5$ and only $\emph{temp}$ of $hws$ is updated. In the third iteration, since the temperature is between $18.5$ and $19$, the guard can be triggered. There are two possible next states; in one state $\emph{hws}$ still remains in $\emph{off}$ mode (state $s_2$) and in another state $\emph{hws}$ switches to the $\emph{on}$ mode (state $S_3$).

In the state $s_3$, a $\emph{control}$ message is sent by $\emph{hws}$ to $c$ with no communication and it is handled in this state. As the value of $\emph{temp}$ is not less than $18.5$, no message is sent. The next state of $s_3$ is achieved by advancing time until $\Delta$ and not shown in the figure. Upon reachability analysis of state $s_2$, the global time interval is advanced for $0.5$. There is a non-determinism; either $\emph{hws}$ stays at $\emph{off}$ (reaching to the state $s_2'$) or switches to $\emph{on}$ (state $s_4$). In the state $s_4$, temperature is definitely less than $18.5$, so a $\emph{notify}$ message is sent with a delay of $[0.3,0.5]$.  Due to the line $12$, we compute the values of $\emph{temp}$ approximately for the time bound $1.5$, i.e., for three stepsizes. As the arrival time of $\emph{notify}$ message is $[1.8,2.8]$, the global time is advanced for $0.3$. There is a non-deterministic behavior in the state $s_4$; either the $\emph{notify}$ message is handled (state $s_5$) or its processing is postponed (state $s_6$). In the state $s_5$, as the global time is advanced for $0.3$, so we have been stayed in the mode $\emph{on}$ for $0.3$. So, the value of $\emph{temp}$ is approximated by the first stored stepsize, resulting $(18,19]$.

\begin{figure}[thpb]
	\centering
	\includegraphics[width=0.85\textwidth]{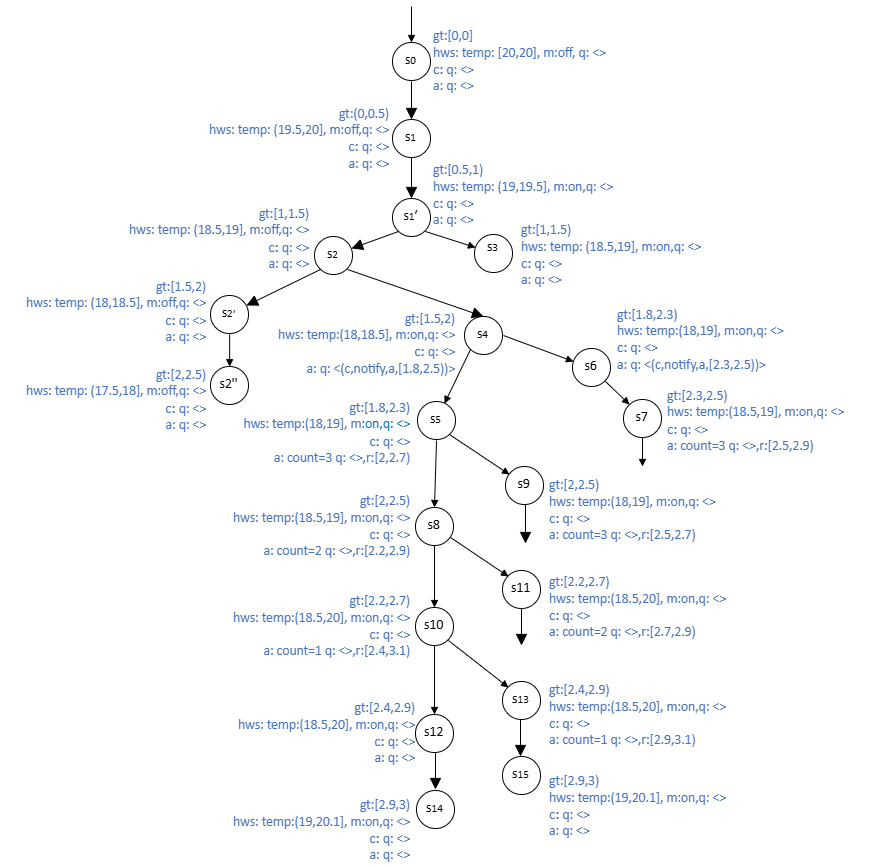}
	\caption{The resulting state space of our reachability algorithm to the Rebeca model in Fig  \ref{code::motiExample}.
	\label{Fig::reachalgo}}
\end{figure}

\end{example}

The advantage of the reachability algorithm in comparison with using $\it TrigInterval$ in cases that ODEs can be solved is that the effect of physical part on the reactive part is considered more precisely. A reactive rebec may behave based on the values it receives from the physical part. The conditional statement in reactive rebec may be inspected for a larger interval (the intervals that a guard may trigger). For example the execution of the conditional statement in state $s_2$ results in a non-deterministic behavior (leading to states $s_5$ and $s_6$) in Figure \ref{Fig::codeTTS}, since the global time is $[1,2)$ and the temperature is $(18,19]$ (this range satisfies and dis-satisfies the condition). However, this non-determinism is resolved in states $s_1'$ and $s_2$ in Figure \ref{Fig::reachalgo}.

\subsection{Soundness of Approach\label{sec::soundness}}
We discuss that the reachable states resulted by our reachability algorithm are subsumed by the reachable states found by executing the reachability algorithm of the corresponding monolithic hybrid automaton. 

\begin{theorem}\label{the::soundness}
For a given Hybrid Rebeca model $\mathcal{M}$, let ${\it HA}(\mathcal{M})$ denote the corresponding monolithic hybrid automaton as defined in \cite{SoSym}. For any reachable state $s$  achieved by Algorithm \ref{alg::HybridRebecaReachabilityAnalysisAlgorithm} on ${\it TTS}(\mathcal{M})$, denoted by $s\in\mathcal{R}_{2}({\it TTS}(\mathcal{M}))$, there exists a reachable state $\langle l,V\rangle$ resulted by Algorithm \ref{alg::Chen} on $HA(\mathcal{M})$, shown by $\langle l,V\rangle\in\mathcal{R}_{1}({\it HA}(\mathcal{M}))$, such that 
$\forall v\in\Var\cdot {\it vals}(s.PS)(v)\subseteq V(v)$, where ${\it vals}(s.PS)\in\Valuation$ is the valuation of physical rebec's variables.
\end{theorem}

\begin{ProofSketch}
We provide a mapping between the reachable states of ${\it TTS}(\mathcal{M})$ and $HA(\mathcal{M})$. Then, we show that each transition of a state of ${\it TTS}(\mathcal{M})$ is mimicked by its mapped state of $HA(\mathcal{M})$ for a the same time interval. As the intervals in the reachability algorithm of ${\it TTS}(\mathcal{M})$ are smaller than the ones in Algorithm \ref{alg::Chen}, more precise values are computed. The proof is provided in  \ref{sec::appendix}.  
\end{ProofSketch}

\subsection{Tool Support\label{subsec::tool}}
Timed Rebeca is supported by Afra, which handles time as a discrete value and the time-related statements (\texttt{delay}(v) and \texttt{after}(v)); these statements have only one discrete parameter. Upon executing the delay statement, the resume time is updated, while upon executing send with an after construct, the arrival time of the message is updated. However, in Hybrid Rebeca, time is continuous while these statements have two parameters imposing non-deterministic behavior as often observed in CPSs. 
%

Due to time non-determinism, global time together with all real variables are maintained as real intervals. The state-space generator of Hybrid Rebeca\footnote{Available at \url{https://github.com/SaeedZhiany/HybridRebecaReachabilityAnalysis}}, implementing Algorithm \ref{alg::HybridRebecaReachabilityAnalysisAlgorithm}, deviates from Afra and therefore was developed from scratch. The tool employs \hypro \cite{hyproToolPaper}, which uses a similar algorithm to \flowStar, based on Taylor model approximation, 
to compute the flowpipes of real variables, i.e., ${\it ComputeFlowpipes}$. We introduce \hypro, and the state-space generator design in \ref{sec::tool}.
Figure \ref{Fig::reachalgo} illustrates the state space obtained from our implementation. We have conducted a set of experiments\footnote{The Hybrid Rebeca models and the tool results are available at \url{https://github.com/SaeedZhiany/HybridRebecaReachabilityAnalysis/tree/feature/duplicate-state/caseStudy}} on our running example with our tool to measure the number of generated states and the tool execution time for different time bounds. We have also extended the running example by adding another sensor; the controller aims to keep the temperature of both sensors above $18.5^\circ$. We have demonstrated the results in Table \ref{Tab::results}.

\begin{table}[h]
\centering
	\caption{Experimental results of applying our tool implementation \label{Tab::results}}
	\begin{tabular}{|cccc|}
		\hline
		~~~No. sensors~~~ & ~~~Time bound~~~ & ~~~No. states~~~ & ~~~Execution time~~~ \\
		\hline
		1 & 1.5s & 6  & 1.18s \\
		1 & 3s & 64  &  1.59s\\		
		1 & 8s & 27989 &  62.78s \\	
		2 & 3s & 81955 &  92.97s \\
		\hline
	\end{tabular}
\end{table}

We remark that our implementation is prototypical, with the aim to inspect the feasibility of our approach. There are several ways in which the implementation could be improved. Such improvements and the examination of scalability aspects is content of future work.




\section{Related Work}\label{sec::related}

\noindent There are different high-level modeling language and tools for specifying hybrid systems. Each employs a specific analysis approach based on the restrictions imposed on their models. 

\noindent \textbf{Theorem prover}: In \cite{platzer2010logical}, a hybrid systems is specified by a set of pre- and post conditions and the property is verified by a theorem prover for time-deterministic models for unbounded time. Hybrid Active Object \cite{kamburjan2019modeling} is an actor-based modeling language. For analysis, a hybrid active object model is converted into a set of pre-  and post-condition in $D\mathcal{L}$ logic \cite{platzer2012complete,platzer2017complete,platzer15} which is fed into \keymaerax tool set for verification \cite{fulton2015keymaera}.\\
\noindent \textbf{Simulation}: is a widely used approach to detect unsafe behavior of hybrid systems \cite{girard2006verification,donze2007systematic,donze2007trajectory}. Hybrid Theatre \cite{nigro2019statistical} is an actor-based language that its analysis is based on statistical model checking of \texttt{UPPAAL SMC} tool set \cite{david2015uppaal} by converting their models to statistical timed or hybrid automata. The modeling language CHARON \cite{alur2000modular} is a hierarchy language for specifying hybrid systems. They provide a modular simulator to simulate each active mode independently. The language CIF \cite{beohar2010hierarchical,agut2013syntax}, based on hybrid automata, supports all types of ODEs by providing a simulation based approach for them. The actor-based modeling framework Ptolemy is a framework that uses the concept of \textit{model of computation} (MoC) which defines the rules for concurrent execution of components and their communications. It is useful for modeling heterogeneous models like process networks, discrete events, dataflow and continuous time, by nesting these models of computations in a hierarchical structure~\cite{ptolemaeus2014system}. \\ 
\noindent \textbf{Reachability Analysis}: The \spaceex toolset \cite{spaceex}, based on hybrid automata, implements a set of reachability analysis algorithms such as \phaver \cite{frehse2008phaver} for linear hybrid systems, Le Guernic-Girard (LGG) \cite{le2010reachability} for non-linear hybrid systems, and efficient LGG \cite{frehse2013flowpipe}. Hybrid Rebeca \cite{jahandideh2018hybrid} relies on \spaceex for the analysis of timed properties on hybrid automata derived from its models. The \flowStar toolset \cite{Chen,flow} has been developed to over-approximate the reachable states of hybrid automata mit non-linear ODEs.

Timed Rebeca is extended with time intervals  to support non-deterministic time behavior in \cite{tavassoli2020finite}. Physical rebecs distinguish our work from it. In other words, our semantic model coincides with their semantics if no physical rebec is specified. Due to the presence of physical rebec, not only global time but also real- and float variables are associated with real intervals. 

\section{Conclusion and Future Work}\label{sec::con}
In this paper, we extended Hybrid Rebeca to be able to model non-deterministic delays for both communication and computation. For their reachability analysis, we adapted an existing algorithm for hybrid automata to be directly applicable to Hybrid Rebeca models, proved the soundness of the method, and illustrated its applicability on a case study.

In future work, 
we aim to improve also the tool implementation, for example by computing the behavior of the conditional statements more precisely based on the intersection of floating-point values and the conditional expressions using the \hypro~tool. Evaluating the precision of our approach or its scalability is also among our future work. 

If a Hybrid Rebeca model has no physical part, then the time-step size can be adjusted to a large value allowing its analysis to be conducted based on the occurrences of events. However, when there is a physical part specified by ODEs, then the time-step size should be adjusted to an appropriately small value. Thus, we also aim at a dynamic adjustment of the time-step size, which we expect to significantly reduce the computational effort. 

We are also planning on conducting more experiments on real-world benchmarks in order to further examine the applicability of our approach.

\bibliographystyle{plain}
\bibliography{references}

\begin{thebibliography}{10}

\bibitem{agut2013syntax}
D.~Agut, D.~van Beek, and J.E Rooda.
\newblock Syntax and semantics of the compositional interchange format for hybrid systems.
\newblock {\em The Journal of Logic and Algebraic Programming}, 82(1):1--52, 2013.

\bibitem{alur2000modular}
R.~Alur, R.~Grosu, Y.~Hur, V.~Kumar, and I.~Lee.
\newblock Modular specification of hybrid systems in {CHARON}.
\newblock In {\em Proc. of the 3rd Int. Workshop on Hybrid Systems: Computation and Control (HSCC'00)}, pages 6--19. Springer, 2000.

\bibitem{alur1995algorithmic}
Ra.v Alur, C.~Courcoubetis, N.~Halbwachs, T.s~A. Henzinger, P.~Ho, X.r Nicollin, A.o Olivero, J.~Sifakis, and S.~Yovine.
\newblock The algorithmic analysis of hybrid systems.
\newblock {\em Theoretical Computer Science}, 138(1):3--34, 1995.

\bibitem{HyCreate}
Stanley Bak and Marco Caccamo.
\newblock Computing reachability for nonlinear systems with {HyCreate}, 2013.

\bibitem{beohar2010hierarchical}
H.~Beohar, D.~Agut, D.~A van Beek, and P.~Cuijpers.
\newblock Hierarchical states in the compositional interchange format.
\newblock {\em arXiv preprint arXiv:1008.2110}, 2010.

\bibitem{julia}
Sergiy Bogomolov, Marcelo Forets, Goran Frehse, Kostiantyn Potomkin, and Christian Schilling.
\newblock {JuliaReach}: {A} toolbox for set-based reachability.
\newblock In {\em Proc. of the 22nd ACM Int. Conf. on Hybrid Systems: Computation and Control (HSCC'19)}, pages 39--44, 2019.

\bibitem{flow}
X.~Chen, E.~{\'{A}}brah{\'{a}}m, and S.~Sankaranarayanan.
\newblock Flow*: {A}n analyzer for non-linear hybrid systems.
\newblock In {\em Proc. of the 25th Int. Conf. on Computer Aided Verification (CAV'13)}, volume 8044 of {\em LNCS}, pages 258--263. Springer, 2013.

\bibitem{Chen}
Xin Chen.
\newblock {\em Reachability analysis of non-linear hybrid systems using {T}aylor Models}.
\newblock PhD thesis, RWTH Aachen University, Germany, 2015.

\bibitem{david2015uppaal}
A.~David, K.~G Larsen, A.~Legay, M.~Miku{\v{c}}ionis, and D.~B. Poulsen.
\newblock Uppaal smc tutorial.
\newblock {\em Int. Journal on Software Tools for Technology Transfer}, 17(4):397--415, 2015.

\bibitem{donze2007trajectory}
A.~Donz{\'e}.
\newblock {\em Trajectory-based verification and controller Synthesis for continuous and hybrid systems}.
\newblock PhD thesis, Universit{\'e} Joseph-Fourier-Grenoble I, 2007.

\bibitem{donze2007systematic}
A.~Donz{\'e} and O.~Maler.
\newblock Systematic simulation using sensitivity analysis.
\newblock In {\em Proc. of the 10th Int. Workshop on Hybrid Systems: Computation and Control (HSCC'07)}, pages 174--189. Springer, 2007.

\bibitem{sapo}
Tommaso Dreossi.
\newblock Sapo: {R}eachability computation and parameter synthesis of polynomial dynamical systems.
\newblock {\em CoRR}, abs/1607.02200, 2016.

\bibitem{frehse2008phaver}
G.~Frehse.
\newblock {PHAVer}: {A}lgorithmic verification of hybrid systems past {HyTech}.
\newblock {\em Int. Journal on Software Tools for Technology Transfer}, 10(3):263--279, 2008.

\bibitem{frehse2013flowpipe}
G.~Frehse, R.~Kateja, and C.~Le~Guernic.
\newblock Flowpipe approximation and clustering in space-time.
\newblock In {\em Proc. of the 16th Int. Conf. on Hybrid Systems: Computation and Control (HSCC'13)}, pages 203--212, 2013.

\bibitem{spaceex}
Goran Frehse, Colas~Le Guernic, Alexandre Donz{\'{e}}, Scott Cotton, Rajarshi Ray, Olivier Lebeltel, Rodolfo Ripado, Antoine Girard, Thao Dang, and Oded Maler.
\newblock {SpaceEx}: {S}calable verification of hybrid systems.
\newblock In {\em Proc. of the 23rd Int. Conf. on Computer Aided Verification}, volume 6806 of {\em LNCS}, pages 379--395. Springer, 2011.

\bibitem{fulton2015keymaera}
Nathan Fulton, Stefan Mitsch, Jan{-}David Quesel, Marcus V{\"{o}}lp, and Andr{\'{e}} Platzer.
\newblock {KeYmaera X}: {A}n axiomatic tactical theorem prover for hybrid systems.
\newblock In {\em Proc. of the 25th Int. Conf. on Automated Deduction (CADE-25)}, pages 527--538. Springer, 2015.

\bibitem{girard2006verification}
Antoine Girard and George~J Pappas.
\newblock Verification using simulation.
\newblock In {\em Proc. of the 9th Int. Workshop on Hybrid Systems: Computation and Control (HSCC'06)}, pages 272--286. Springer, 2006.

\bibitem{henzinger2000theory}
T.~A. Henzinger.
\newblock The theory of hybrid automata.
\newblock In {\em Proc. of the 11th Annual IEEE Symposium on Logic in Computer Science (LICS'96)}, pages 278--292. IEEE, 1996.

\bibitem{SoSym}
I.~Jahandideh, F.~Ghassemi, and M.~Sirjani.
\newblock An actor-based framework for asynchronous event-based cyber-physical systems.
\newblock {\em Software and Systems Modeling}, 20(3):641--665, 2021.

\bibitem{jahandideh2018hybrid}
Iman Jahandideh, Fatemeh Ghassemi, and Marjan Sirjani.
\newblock Hybrid {R}ebeca: {M}odeling and analyzing of cyber-physical systems.
\newblock In {\em Proc. of the 8th Int. Workshop on Model-Based Design of Cyber Physical Systems (CyPhy'18)}, volume 11615 of {\em LNCS}, pages 3--27. Springer, 2018.

\bibitem{kamburjan2019modeling}
E.~Kamburjan, S.~Mitsch, M.a Kettenbach, and R.r H{\"a}hnle.
\newblock Modeling and verifying cyber-physical systems with {Hybrid Active Objects}.
\newblock {\em arXiv preprint arXiv:1906.05704}, 2019.

\bibitem{nigro2019statistical}
L.~Nigro and P.~F. Sciammarella.
\newblock Statistical model checking of cyber-physical systems using {Hybrid Theatre}.
\newblock In {\em Proc. of the 2019 Conf. on Intelligent Systems and Applications (IntelliSys'19)}, pages 1232--1251. Springer, 2019.

\bibitem{platzer2012complete}
A.~Platzer.
\newblock The complete proof theory of hybrid systems.
\newblock In {\em Proc. of the 27th Annual IEEE Symp. on Logic in Computer Science (LICS'12)}, pages 541--550. IEEE, 2012.

\bibitem{platzer2017complete}
A.~Platzer.
\newblock A complete uniform substitution calculus for differential dynamic logic.
\newblock {\em Journal of Automated Reasoning}, 59(2):219--265, 2017.

\bibitem{platzer15}
A.~Platzer.
\newblock {\em Logical Foundations of Cyber-Physical Systems}.
\newblock Springer, 2018.

\bibitem{platzer2010logical}
Andr{\'{e}} Platzer.
\newblock {\em Logical Analysis of Hybrid Systems - Proving Theorems for Complex Dynamics}.
\newblock Springer Science \& Business Media, 2010.

\bibitem{ptolemaeus2014system}
C.~Ptolemaeus.
\newblock {\em System design, modeling, and simulation using {Ptolemy II}}, volume~1.
\newblock Ptolemy. org Berkeley, 2014.

\bibitem{aceto2011modelling}
A.~Hermann Reynisson, M.~Sirjani, L.~Aceto, M.~Cimini, A.~Jafari, A.~Ing{\'{o}}lfsd{\'{o}}ttir, and S.~Hugi Sigurdarson.
\newblock Modelling and simulation of asynchronous real-time systems using {Timed Rebeca}.
\newblock {\em Science of Computer Programming}, 89:41--68, 2014.

\bibitem{hyproToolPaper}
Stefan Schupp, Erika {\'A}brah{\'a}m, Ibtissem~Ben Makhlouf, and Stefan Kowalewski.
\newblock {HyPro}: {A} {C}++ library of state set representations for hybrid systems reachability analysis.
\newblock In Clark Barrett, Misty Davies, and Temesghen Kahsai, editors, {\em Proc. of the 9th Int. Symp. on NASA Formal Methods (NFM'17)}, pages 288--294, Cham, 2017.

\bibitem{sirjani2004modeling}
M.~Sirjani, A.~Movaghar, A.~Shali, and F.~S. de~Boer.
\newblock Modeling and verification of reactive systems using {R}ebeca.
\newblock {\em Fundamenta Informaticae}, 63(4):385--410, 2004.

\bibitem{tavassoli2020finite}
Shaghayegh Tavassoli, Ramtin Khosravi, and Ehsan Khamespanah.
\newblock Finite interval-time transition system for real-time actors.
\newblock In {\em Proc. of the 3rd Int. Conf. on Topics in Theoretical Computer Science (TTCS'20)}, pages 85--100. Springer, 2020.

\bibitem{le2010reachability}
Mark Wetzlinger, Niklas Kochdumper, Stanley Bak, and Matthias Althoff.
\newblock Fully-automated verification of linear systems using reachability analysis with support functions.
\newblock In {\em Proc. of the 26th ACM Int. Conf. on Hybrid Systems: Computation and Control (HSCC'23)}, pages 5:1--5:12. ACM, 2023.

\bibitem{Zha92}
F.~Zhao.
\newblock {\em Automatic Analysis and Synthesis of Controllers for Dynamical Systems Based on Phase-Space Knowledge}.
\newblock Ph.d, Massachusetts Institute of Technology, 1992.

\end{thebibliography}
\appendix 
\section{Proof of Theorem \ref{the::soundness}\label{sec::appendix}}

We first explain the rules deriving a monolithic hybrid automaton as the semantic model of a given Hybrid Rebeca model. Next, we provide the rules using the semantic rules of both semantics.

\subsection{Semantics of Hybrid Rebeca models based on hybrid automata}
The syntax of Hybrid Rebeca model introduced in \cite{SoSym} contains the network specification that shows the communication delays among the rebecs. The syntax of the Hybrid Rebeca model has no network specification, and instead communication delays are addressed by the $\mathsf{after(C,c)}$ statement and the send message statement. We adapt the semantic model proposed in \cite{SoSym} to support the syntax in this paper. 

\begin{definition} [Hybrid automaton for a Hybrid Rebeca model]
Given a Hybrid Rebeca model  $\mathfrak{H}=(R_s,R_p,N)$, its formal semantics are based on hybrid automata which is defined as ${\it HA}(\mathcal{M})=(\Loc,\Var,\Jumps, \Flws, \Inv, \Init)$, where $\Var$ is the set of all continuous variables in the model (variables of types \textit{float} or \textit{real}), transitions $\Jumps$, flows $\Flws$, invariants $\Inv$, and initial conditions $\Init$ are defined below.
\end{definition}


\subsubsection{Locations}
Each location is defined by four entities, denoted by the record  $(\DS:\ds,\CS:\cs,\ES:\es)$. The entity $\ds$ defines the states of reactive rebecs by mapping a given reactive rebec with the identifier $x$ to its local state. Similarly $\cs$ maps a physical rebec with the identifier $x$  
to its local state. The third entity $\es$ represents the sequence of pending events. The local states of reactive rebecs are defined by a tuple $(\sigma,q,st,r)\in \Valuation \times \Msg^* \times \Stmt^* \times \Interval \times \{\top,\bot\}$, where $\valuation $ is the valuation function that maps only int variables to their value, $q$ is the rebec's mailbox, $st \in \Stmt^*$ is the sequence of statements of current message server (the rebec must execute them) and $r$ denotes where the rebec is suspended ($\top$) or not ($\bot$). The messages in the mailbox are the same as in TTS but with no arrival time. The  local states of physical rebecs are defined by the tuple $(q,st,\m) \in \Msg^* \times \Stmt^*  \times \modes$  where $q$ is the rebec's mailbox, $st \in \Stmt^*$ is the sequence of statements of current message server (the rebec must execute them) and  $\m$ donates the rebec's current active mode.

Pending events are used for time consuming actions: executing delay statements or transferring messages with a delivery delay. Upon executing a time consuming action, a pending event is stored in $\es$ to be triggered at the time that the delay of the action is over. Two types of pending events are defined in the semantics of Hybrid Rebeca: \EventResume\ and \EventEnqueue. Let ${\it PEvent}$ denote the set of all pending events, specified by $\{\EventEnqueue(x,m),\EventResume(x)\,\mid\, x\in\ID 
,\, m\in {\it Msg}\}$. A pending event with the event \EventResume, parameterized by a rebec identifier, is generated and inserted into the pending event list when a delay statement is executed in the Rebeca model, and the corresponding rebec is suspended. To model the passage of time for the delay statement, a timer is assigned to the pending event. After the specified delay has passed, the pending event is triggered, and consequently the behavior of the given rebec is resumed by updating the suspension status of the rebec. 
The pending event $\EventEnqueue(x,m)$ is generated when a message with a delivery delay is chosen to be delivered to its receiver. A timer is assigned to model the message delivery delay, and the pending event is inserted into the pending event list. Upon triggering a \EventEnqueue event, the specified message is enqueued in the receiver's message queue.  The effect of each pending event trigger on a given location is formally defined by a function $\EventEffectFunction: {\it Event}\times \Loc \rightarrow \Loc $ which is presented in Table \ref{Tab:EventDefs}.

\begin{table}[htbp]
	\centering
\caption{Formal definition of the function $\EventEffectFunction$}
\label{Tab:EventDefs}
	
	\begin{tabular}{rl}
		
		{\EventResume} & $\Rule{\gsDef \wedge \ds(x)=(\sigma,q,st,r)}
		{\EventEffectFunction(\EventResume(x),\gs)=  \gs[\ds[x \mapsto(\sigma,q,st,\bot)]/\DS]}$\vspace*{3mm}
		\\
		{\EventEnqueue (Software)} & $\Rule{\gsDef \wedge \ds(x)=(\sigma,q,st,r) 
        }
		{\EventEffectFunction(\EventEnqueue(x,m),\gs)=  \gs[\ds[x \mapsto(\sigma,q\frown m,st,r)]/\DS]}$
		\vspace*{3mm}\\
		{\EventEnqueue (Physical)} & $\Rule{\gsDef \wedge \cs(x)=(q,st,\mathfrak{m})}
		{\EventEffectFunction(\EventEnqueue(x,m),\gs)=  \gs[\cs[x \mapsto(q \frown m,st,\mathfrak{m})]/\CS}$\vspace*{3mm}
	\end{tabular}
\end{table}

\begin{definition}[Pending event]
	A pending event is a tuple $(d_1,d_2,e,t)$ where $d_1$ and $d_2$ specify the interval delay for event $e$ and $t$ is a timer  that is assigned to this event. The event $e$ can either be a \EventResume\ or \EventEnqueue\ event. The timer is a real valued variable used for defining the timing behavior for the delay of the pending event. The event can be triggered (and executed) in a time $d_1 \le t\le d_2$ after the pending event is created.
	
\end{definition}

\subsubsection{Transitions}
There are two types of transitions: urgent and non-urgent transitions to differentiate between different types of actions. An urgent transition, shown by $\Rightarrow_{\UrgentTranSymbol}$, must be taken immediately upon entering its source location. 
The non-urgent transitions are shown as \NonurgentTransition. These transitions indicate the passage of time.
These transitions include the behaviors of physical rebecs' active modes and pending time of events since they are time consuming. 
Urgent transitions have a higher priority than the non-urgent ones;  a non-urgent transition is enabled if no urgent transition is enabled. 

\def\rulespace{3mm}

\subsubsection*{Urgent Transitions:} 
Urgent transitions are taking a message from the mailbox or execute a statement. 

\begin{itemize}
\item \textit{Taking a message}: 

\begin{center}	
$\Rule
{\gsDef \wedge \ds(x)=(\sigma,q,\epsilon,\bot)\wedge m\in q }
{\gs\xRightarrow[]{\tau}_{\UrgentTranSymbol}\gs[\ds[x\mapsto (\sigma,q,body(x,m),\bot)]/\DS]
}$ \vspace*{\rulespace}

	\end{center}	


\item \textit{Assignment Statement:} There are two rules for the assignment statement based on the type of the left variable: assigning to a discrete variable $\it dvar$ and assigning to a continuous variable $\it cvar$. Since the right value for the continuous case is not determined, the assignment is transferred over to the transition to be handled by the resulting hybrid automaton.	
	\begin{center}	
		$\Rule
		{ \gsDef \wedge	\ds(x)=(\sigma,q,\langle dvar=e|st\rangle,\bot)  }
		{\gs\xRightarrow[]{\tau}_{\UrgentTranSymbol} \gs[\ds[x \mapsto(\sigma[dvar\mapsto eval(e,v)],q,st,\bot)]/\DS]  
		}$ \vspace*{\rulespace}
	
		$\Rule
		{ \gsDef \wedge	\cs(x) = (q, \langle cvar=e|st\rangle,\mathfrak{m})
			}
		{\gs\xRightarrow[]{cvar:=eval(e,v)}_{\UrgentTranSymbol} \gs[\cs[x \mapsto (q,st,\mathfrak{m})]/\CS] 
		}$
	\end{center}
	\item \textit{Conditional Statement:} This statement has three rules depending on the value of the condition. 
    The value of the the condition may not be determined (because of continuous variables used in the condition). So, both possible paths are considered by creating two separate transitions. The condition and its negation act as the guards for these transitions. 
	\begin{center}	
	$\Rule{\gsDef \wedge	\ds(x)=(\sigma,q,\langle \text{if}~(e)~\text{then}~st_1~\text{else}~st_2 |st''\rangle,\bot) \wedge 	eval(e,v) = {\it true}  }
	{\gs\xRightarrow[]{\tau}_{\UrgentTranSymbol} \gs[\ds[x \mapsto(\sigma,q,st_1 \frown st '',\bot)]/\DS] 
    }$ \vspace*{\rulespace}
    
	$\Rule
	{\gsDef \wedge	\ds(x)=(\sigma,q,\langle \text{if}~(e)~\text{then}~st_1~\text{else}~st_2 |st''\rangle,\bot) \wedge eval(e,v) = {\it false}}
	{\gs\xRightarrow[]{\tau}_{\UrgentTranSymbol} \gs[\ds[x \mapsto(\sigma,q,st_2 \frown st'',\bot)]/\DS] 
		}$\vspace*{\rulespace}
        
	$\Rule
	{\gsDef \wedge	\cs(x)=(\langle \text{if}~(e)~\text{then}~st_1~\text{else}~st_2 |st''\rangle,\mathfrak{m})
	}
	{
    \begin{array}{c}
	\gs\xRightarrow[]{e}_{\UrgentTranSymbol} \gs[\cs[x \mapsto(q,st_1 \frown st'',\mathfrak{m})]/\CS] \\
	\gs\xRightarrow[]{!e}_{\UrgentTranSymbol} \gs[\cs[x \mapsto(q,st_2 \frown st'',\mathfrak{m})]/\CS]  
    \end{array}
	}$
\end{center}

\item \textit{Send Statement:} This statement, depending on the delivery delay has two rules. 
The same rules hold for physical rebecs.
\begin{center}	
 $\Rule
  {\gsDef \wedge	\ds(x)=(\sigma_x,q_x,\langle(y.m()) |st_x\rangle,\bot)  \wedge
  \ds(y) = (\sigma_y,q_y,st_y,r) 
  }
  {\gs\xRightarrow[]{\tau}_{\UrgentTranSymbol} \gs[\ds[x \mapsto(\sigma_x,q_x,st_x,\bot),y\mapsto(\sigma_y,q_y \frown (x,m,y),st_y,r)]/\DS] }$\vspace*{\rulespace}
 $\Rule
 {\gsDef \wedge	\ds(x)=(\sigma,q,y.m()~ \mathsf{after}(c_1,c_2) |st\rangle,\bot) 
}
{\gs\xRightarrow[]{\tau}_{\UrgentTranSymbol} \gs[\ds[x \mapsto(\sigma,q,st,\bot)]/\DS,\es \frown (c_1,c_2,\EventEnqueue(y,(x,m,y)),t)/\ES 
]}$
\end{center}

\item \textit{Delay Statement:} This statement suspends the software rebec and creates a pending event $(c_1,c_2,\EventResume,t)$ for resuming the rebec within the interval $[c_1,c_2)$. 
The timer $t$ is a fresh timer acquired from the pool of timers by calling $acquire\_timer()$.
\begin{center}	
   $\Rule
	{\gsDef \wedge \ds(x)=(\sigma,q,\langle delay(c_1,c_2)|st\rangle,\bot) \wedge 
	t = \textit{acquire\_timer()}}
	{
    \gs\xRightarrow[]{\tau}_{\UrgentTranSymbol}\gs[ \ds[x\mapsto(\sigma,q,st,\top)]/\DS, \es \frown (c_1,c_2,\EventResume(x),t)/\ES]
	}$
	\end{center}

	\item \textit{Set Mode Statement:} 
	\begin{center}
	$\Rule
	{	\gsDef \wedge \cs(x)=(q,\langle {\it setmode}(\mathfrak{m}'|st\rangle,\mathfrak{m} )}
	{\gs\xRightarrow[]{\tau}_{\UrgentTranSymbol} \gs[\cs[x\mapsto (q,st,\mathfrak{m}')]/\CS]}
    $
	\end{center}

\end{itemize}

\subsubsection*{Non-urgent Transitions:} 
Non-urgent transitions are used to define the end of active physical modes and trigger pending events: 
\begin{itemize}
	\item \textit{End of an Active Mode:} Let $guard(x,\mathfrak{m})$ and ${\it Trigger}(x,\mathfrak{m})$ denote the guard and trigger for the mode $\mathfrak{m}$ in the class of rebec $x$. 
\begin{center}
	$\Rule{\gsDef \,\wedge\, \cs(x)=( \epsilon, \epsilon,\mathfrak{m}) \,\wedge\, \mathfrak{m} \neq {\it none} \wedge 
	\NotUrgentTransition }
	{\gs \xRightarrow{guard(x,\mathfrak{m})}_{\NonurgentTranSymbol} \gs[\cs[x\mapsto (\epsilon, {\it Trigger}(x,\mathfrak{m}),{\it none})]/\CS]
		}$
	\end{center}
	
	\item \textit{Triggering of an Event:} For a pending event  $(d_1,d_2,e,t)$, the guard $(d_1\le t)$  is defined on the transition where $t$ is the timer. 
    The event $e$ is executed as a result of this transition and the pending event is removed from the pending event list. The effects of the execution of an event is defined by \EventEffectFunction\ function given in Table \ref{Tab:EventDefs}.
	\begin{center}	
	$\Rule
	{\gsDef \,\wedge\,  (d_1,d_2,ev,t)\in \es\,\wedge\,
	\NotUrgentTransition}
	{\gs\xRightarrow{(d_1\le t)}_{\NonurgentTranSymbol} {\it \EventEffectFunction}(ev,\gs[\es\setminus (d_1,d_2,ev,t)/\ES])
		}$
	\end{center}  
\end{itemize}

\subsubsection{Flows and Invariants}
The flows and invariants for each location are defined based on its urgent and non-urgent transitions. Urgent transitions should be executed without allowing the time passage. Meaning that time should not progress when the system is residing in the source locations of an \emph{urgent} transition. Any location with an urgent transition is considered an urgent location.
\begin{definition}[Urgent location flow and invariant]\label{Def::urgent}
A possible implementation for an urgent location is ${\it urg}'=1$ as its flow and ${\it urg}\le 0$ as its invariant, where  ${\it urg}$ is a specific variable. Note that in this method, this new variable must be added to the set $\Var$ of the hybrid automaton. Also the assignment ${\it urg} = 0$ must be added to all incoming transitions to an urgent location as the reset of the transition. The defined invariant prevents the model from staying in the location as the value of $\it urg$ will be increased by the defined flow. 
\end{definition}

If a location is urgent, the urgency flow, as defined above, should be set as its flows. In case a location is not urgent, it inherits the flows of all physical rebecs' active modes, denoted by \ModeFlows, the flows for timers of pending events, denoted by \EventFlows, and a flow of zero for each float variable to freeze its value, denoted by \ConstantFlows. The flow of a pending event is simply defined as $t'=1$ where $t$ is the timer variable of the pending event. The \Flws  function of the hybrid automaton is defined as:

\[\Flws(\gs)=\begin{cases}
\UrgFlow,& \text{if $\gs$  is urgent} \\
\ModeFlows(\gs) \bigcup \EventFlows(\gs) \bigcup \ConstantFlows(\gs) ,& \text{otherwise}
\end{cases}\]

\[ \begin{array}{ll}
\ModeFlows(\gs)  &= \sideset{}{}\bigcup_{x \in R_p} {\it flows}(x,\mathfrak{m})\ \text{where}\ \gs.\CS(x) = (\sigma, q, st,\mathfrak{m}) \\
\EventFlows(\gs) &= \sideset{}{ }\bigcup_{(d, e, t) \in \gs.\ES} t'=1 \\
\ConstantFlows(\gs) &= \sideset{}{}\bigcup_{v \in \text{float variables}} v'=0\ 
\end{array}\]

Similarly, if a location is urgent, its invariant is set to urgency invariant, otherwise it inherits the invariants of all physical rebecs' active modes, denoted by \ModeInvs\, and the invariants for corresponding pending events' timers, denoted by \EventInvs. The invariant of a pending event is defined as $t \le d$ where $t$ and $d$ are the timer variable and the delay of the pending event, respectively. The \Inv\ function of the hybrid automaton is defined as:

\[\Inv(\gs)=\begin{cases}
\UrgInv,& \text{if $\gs$  is urgent} \\
\ModeInvs(l) \wedge \EventInvs(l)  ,& \text{otherwise}
\end{cases}\]

\[ \begin{array}{ll}
\ModeInvs(\gs)  &= \sideset{}{}\bigwedge_{x \in R_p} {\it invariant}(x,\mathfrak{m})\ \text{where}\ \gs.\CS(x) = (\sigma, q, st,\mathfrak{m}) \\
\EventInvs(\gs) &= \sideset{}{ }\bigwedge_{(c_1,c_2, e, t) \in \gs.\ES}t < c_2 \\
\end{array}\] 

\subsubsection{Initial Location and Initial Condition}

The initial location $\gs_0$ is achieved by initializing the state variables of rebecs to their default values, i.e., zero, and their statements to the statement of their constructors. 
The function $\Init$ of the hybrid automaton are defined as:

\[\Init(\gs)=\begin{cases}
\sideset{}{}\bigcup_{v \in \text{continuous variables}} v=0\, & \text{if $\gs = \gs_0$} \\
\emptyset  ,& \text{otherwise}
\end{cases}\]

\subsection{Proof of the Theorem}

We provide a set of lemmas to address the proof of Theorem \ref{the::soundness}.

We revise the SOS rules for the send and delay statements such that the global time that these statements are executed together with the parameters $c_1$ and $c_2$ in the constructs $\mathsf{after}(c_1,c_2)$ and $\mathsf{delay}(c_1,c_2)$ are appended to the arrival time of message or the resume of the rebec. This assumption is helpful for the following definition. So a message $(y,m,p,ar)\in\Msg$, $ar$ has four parts: $ar.t$, $ar.i$, $ar.c_1$, and $ar.c_2$  denoting the arrival time of the message, the insertion time of the message, the lower, and upper bounds of communication delay. For a local state of a reactive rebec $(\sigma, \bag, st,r)$, $r$ has four parts: $r.t$, $r.i$, $r.c_1$, and $r.c_2$  denoting the resume time of the rebec, suspension time of the rebec, the lower, and upper bounds of computation delay. The following definition transforms a state of TTS into a pair of a monolithic hybrid automaton location and a valuation for real/float variables.

\begin{definition}\label{Def::map}
For a given Hybrid Rebeca model $\mathcal{M}$, let $s=(ss,sp,gt)$ be a state of ${\it TTS}(\mathcal{M})$. 
Then $\map(s)=\langle (hs,hp,he),V\rangle$ is defined as\[
\begin{array}{l}
\forall x\in\ID_s\exists \sigma_s,\sigma_h,\bag,q,st,r_s,r_h\cdot ss(x)=(\sigma_s,\bag,st,r_s) \wedge hs(x)=(\sigma_h,q,st,r_h)\,\wedge\\
(\forall v\in\Dom(\sigma_s)\cdot (\Type(x)=\int\Rightarrow 
\sigma_s(v)=\sigma_h(v) )\vee (\Type(v)=\Interval\Rightarrow \sigma_s(v)=V(v))
\, \wedge\\
\forall (y,m,p,ar)\in\bag\cdot ((ar.t_{\it up}< gt_{\it up}\wedge (y,m,p)\in q)\,\vee \\
\hspace*{0.5cm}(
\exists (ar.c_1,ar.c_2,\EventEnqueue(x,(y,m,p)),\mathfrak{t})\in he \wedge V(t)=[\alpha(ar),\beta(ar))))\,\wedge\\
(({r_s}\neq \bot\wedge {r_h}=\top \wedge \exists (r_s.c_1,r_s.c_2,\EventResume(x),\mathfrak{t})\in he \wedge V(t)=[\alpha(r_s),\beta(r_s))\vee
\\
\hspace*{0.5cm}({r_s}= \bot\wedge {r_h}=\bot))\,\wedge\\
\forall x\in\ID_p\exists \sigma_p,\mathfrak{m}\cdot sp(x)=(\sigma_p,\emptyset,\emptyset,\mathfrak{m}) \wedge hp(x)=(\emptyset,\emptyset,\mathfrak{m})\,\wedge\\
(\forall v\in\Dom(\sigma_p)\cdot ( 
\sigma_p(v)=V(v))\,

\end{array}
\] where $\mathfrak{t}$ is a timer, and $\alpha(z)=gt_{\it low}-z.i_{\it low}$ and $\beta(z)=\alpha(z)+(gt_{\it up}-gt_{\it low})$.
\end{definition}

\begin{lemma}\label{Lem::urgentprogress}
For a given Hybrid Rebeca model $\mathcal{M}$, assume $s$ is a state of ${\it TTS}(\mathcal{M})$ 
such that $\map(s)=\langle l,V\rangle$. 
For any transition $s\Rightarrow_{\UrgentTranSymbol}s'$ there exists $l'$ such that $l\Rightarrow^* l'$, $\map(s')=\langle l',V\rangle$, and $\map(s')\in\mathcal{R}_{1}({\it HA}(\mathcal{M}))$.
\end{lemma}
\begin{Proof}
Five cases can be considered based on the SOS rule resulting $s\Rightarrow_{\UrgentTranSymbol}s'$. We only discuss the Take message rule as the other cases can be discussed with a similar fashion. 

The Take message rule implies that there exists the rebec $x$ such that $s.RS(x)=(\sigma,b,\epsilon,\bot)$,  $(s,m,p,r)\in \bag$, and $s'=s\left[rs\left[x\mapsto(\sigma \uplus p,\bag \setminus msg,body(x,m),\bot)\right]/RS\right]$. According to Definition \ref{Def::map} either (1) the message $(s,m,p)$ is in the mailbox of the  rebec $x$ or (2) a $(d_1,d_2,\EventEnqueue(x,m),\mathfrak{t})$ event exists in the pending event list of $\map(s)$. In the first case, $l$ is an urgent location. So $\it ComputeFlowpipes$ in line $8$ of Algorithm \ref{alg::Chen} only returns $V$. By taking the message SOS rule, $l\xRightarrow[]{\tau}_{\UrgentTranSymbol}l'$, where $l'=l[\ds[x\mapsto (\sigma,q,body(x,m),\bot)]/\DS]$ (lines $9-16$ of Algorithm \ref{alg::Chen}), only the message is removed from the mailbox and the statements of rebec $x$ is updated to $body(x,m)$. Obviously it holds that $\langle l',V\rangle\in\mathcal{R}_{1}({\it HA}(\mathcal{M}))$ and $\map(s')=\langle l',V\rangle$. 

In the second case, by the Triggering event SOS rule, $l\xRightarrow{d_1\le t}_{\NonurgentTranSymbol}l''$, where the message has been inserted into the mailbox of $x$ in $l''$ and the event has been from from the pending event list of $l''$. We remark that  $V\in\it ComputeFlowpipes$ in line $8$, so it  holds that $\langle l'',V\rangle\in\mathcal{R}_{1}({\it HA}(\mathcal{M}))$. The location $l''$ is an urgent location. By the taking the message SOS rule, $l''\xRightarrow[]{\tau}_{\UrgentTranSymbol}l'$ where $l'=l''[\ds[x\mapsto (\sigma,q,body(x,m),\bot)]/\DS]$ (lines $9-16$ of Algorithm \ref{alg::Chen}). Trivially, it holds that $\langle l',V\rangle\in\mathcal{R}_{1}({\it HA}(\mathcal{M}))$ and $\map(s')=\langle l',V\rangle$.

\end{Proof}

\begin{lemma}\label{Lem::non-urgentprogress}
For a given Hybrid Rebeca model $\mathcal{M}$, assume $s$ is a state of ${\it TTS}(\mathcal{M})$. 
For any $ s'\in ,\mathcal{R}_{2}({\it TTS}(\mathcal{M}))$ as a consequence of executing lines $7-28$ of Algorithm \ref{alg::HybridRebecaReachabilityAnalysisAlgorithm} for $s\in \mathcal{R}_{2}({\it TTS}(\mathcal{M}))$, then $\map(s')\in\mathcal{R}_{1}({\it HA}(\mathcal{M}))$ as a consequence of running Algorithm \ref{alg::Chen} upon handling $\map(s)\in\mathcal{R}_{1}({\it HA}(\mathcal{M}))$.
\end{lemma}

\begin{Proof}
Trivially $s\not\nonTimePassingTransition$ as all the non-time progressing transitions have been applied before inserting $s$ into the queue in lines $3$ and $25$. So, the corresponding location of $\map(s)$ is a non-urgent location.

We assume that $(\mathcal{J}>0)\wedge (GT_{\it low}\neq \Delta)$. By calling  ${\it processTime}$ in line $10$ in  Algorithm \ref{alg::HybridRebecaReachabilityAnalysisAlgorithm}, we can consider two cases based on the result  for the new global time, denoted by $gt'$: \begin{itemize}
\item $gt'=[gt_{\it up},t)$ where $t$ can be either the lower bound or upper bound of an event like $e$. The event can be either the arrival of a message or resuming a rebec. First, the values of real variables are updated as long as the invariants of physical rebecs are satisfied (lines $11-12$ in Algorithm \ref{alg::HybridRebecaReachabilityAnalysisAlgorithm}). Let the resulting state be $s''$. In Algorithm \ref{alg::Chen}, in lines $7$ and $8$, the real variables are updated. We remark that the float variables will not change as their flows is $0$ (see Definition \ref{Def::urgent}). So, $\map(s'')\in\mathcal{R}_{1}({\it HA}(\mathcal{M}))$.

By inspecting the guards of physical rebecs's modes in line $17-23$, some of the rebecs may do discrete jump, resulting in a state like $s'$. Assume that the physical rebec $p$ triggers its mode. By the End of an Active mode SOS rule, $l''\xRightarrow{guard(p,\mathfrak{m})}_{\NonurgentTranSymbol}l'$, where $\map(s'')=\langle l'',V''\rangle$ and the trigger's of $p$ is updated in $l'$. Trivially $\map(s')=(l',V'')$ and $\map(s')\in\mathcal{R}_{1}({\it HA}(\mathcal{M}))$. 

Two cases can be distinguished whether $t$ is the lower or upper bound of $e$ :
\begin{enumerate}
\item If $t$ is the lower bound of $e$, then reactive rebecs can not be updated and only the physical rebecs are updated. The only non-time progressing transitions that can be applied to $s'$, belong to the execution of physical rebec's triggers and no reactive rebec is updated in line $25$.  By Lemma \ref{Lem::urgentprogress}, for any $s^\ast$ that $s'\Rightarrow_{\UrgentTranSymbol}^* s^\ast$ and $\map(s')\in\mathcal{R}_{1}({\it HA}(\mathcal{M}))$, then $\map(s^\ast)\in\mathcal{R}_{1}({\it HA}(\mathcal{M}))$. 

\item If $t$ is the upper bound of $e$, so this event can be handled and the reactive part can be updated. If $e$ is the arrival of a message, the message can be handled by the rebec $x$ by applying the Take message SOS rule when ${\it executeNTPSOS}(s')$ is called. According to Definition \ref{Def::map} either (1) the message is in the mailbox of the  rebec $x$ or (2) a $(d_1,d_2,\EventEnqueue(x,m),\mathfrak{t})$ event exists in the pending event list of $\map(s')$. The first case is trivial by lemma \ref{Lem::urgentprogress}. In the second case,  by the Triggering an event SOS rule, $l'\xRightarrow{d_1\le t}_{\NonurgentTranSymbol}l'''$, where the message has been inserted into the mailbox of $x$ in $l'''$ and $(l''',V'')\in\mathcal{R}_{1}({\it HA}(\mathcal{M}))$. 
If $e$ is the resume of a rebec, the rebec is resumed by applying the Resume SOS rule when ${\it executeNTPSOS}(s')$ is called. According to Definition \ref{Def::map} a $(d_1,d_2,\EventResume(x),\mathfrak{t})$ event exists in the pending event list of $\map(s')$. By the Triggering an event SOS rule, $l'\xRightarrow{d_1\le t}_{\NonurgentTranSymbol}l'''$, where the message has been inserted into the mailbox of $x$ in $l'''$ and $(l''',V'')\in\mathcal{R}_{1}({\it HA}(\mathcal{M}))$. 
\end{enumerate}
\item $gt'=[t_1,t_2]$ where $t_1$ and $t_2$ corresponds to the events $e_1$ and $e_2$. Three cases can be distinguished:\begin{enumerate} 
\item If $t_1$ is the upper bound of $e_1$ and $t_2$ is the lower bound of $e_2$, no event can be handled and only the physical part is updated. This can be discussed similar to the case $(1)$ in the previous bullet.

\item If $t_1$ and $t_2$ are both the lower bounds of $e_1$ and $e_2$, only $e_1$ can be handled. If $t_1$ and $t_2$ are the upper bounds of $e_1$ and $e_2$, only $e_2$ can be handled. This can be discussed similar to the case $(2)$ in the previous bullet.
\item If $t_1$ is lower bound of $e_1$ and $t_2$ is the upper bound of $e_2$, both events $e_1$ and $e_2$ can be handled. This case can be discussed by taking two discrete jumps in  ${\it HA}(\mathcal{M})$ while the values of real variables are updated for the once.

\end{enumerate}

\end{itemize}

The proof of Theorem \ref{the::soundness} is concluded from lemmas \ref{Lem::urgentprogress} and \ref{Lem::non-urgentprogress}. 
\end{Proof}

\section{Tool Implementation\label{sec::tool}}
We explain Hypro, and the state-space generator design.
\
\subsection{Hypro}

To calculate the time successor sets from the initial set we used an open-source tool called Hypro \cite{hyproToolPaper}. It provides implementations for the most prominent state set representations used by flowpipe-construction-based reachability analysis techniques for linear hybrid systems and Taylor model based reachability analysis for non-linear hybrid systems.
The only parameters needed to be specified by the user, are the time horizon, jump depth and step size. 

The interaction with Hypro was possible through a Java Native Interface (JNI) type interface, which allows any Java program to invoke time evolution calculations for a given ODE and intersect the resulting reachable sets with a given guard.  

Hypro uses a similar algorithm to \flowStar, based on Taylor model approximation, for calculating time successor sets from initial sets. 
During the analysis, Hypro subdivides the given time horizon $[0, t]$ into time step length sub-intervals, following the flowpipe construction scheme. 
In each sub-interval the activation function is approximated with an at most $\mathit{k} \in \mathbf{N}_{> 0}$ order \emph{Taylor polynomial} and a \emph{remainder interval} $\mathit{I} \subset \R$. A Taylor polynomial is an order $k$ polynomial derived from a functions' derivatives around a given value, according to an infinite sum, called the \emph{Taylor series}. 
Since the Taylor series is an infinite sum, but the Taylor polynomial is a finite approximation of the activation function, the error that arises from the approximation is counted for, in the reminder interval. From the Taylor polynomial and the remainder interval, the reachable states can be derived for each time point in $[0, t]$.

\subsection{State-space Generator}
The semantic model of a given Hybrid Rebeca model is generated by calling the method \texttt{analyzeReachability()} from the $\it SpaceStateGenerator$ class. This method implements the Algorithm \ref{alg::HybridRebecaReachabilityAnalysisAlgorithm}. It initializes the current global time and the parameters for the analysis, e.g., step size and the time-bound. 

 \begin{figure}
	\centering
	\includegraphics[width=\textwidth]{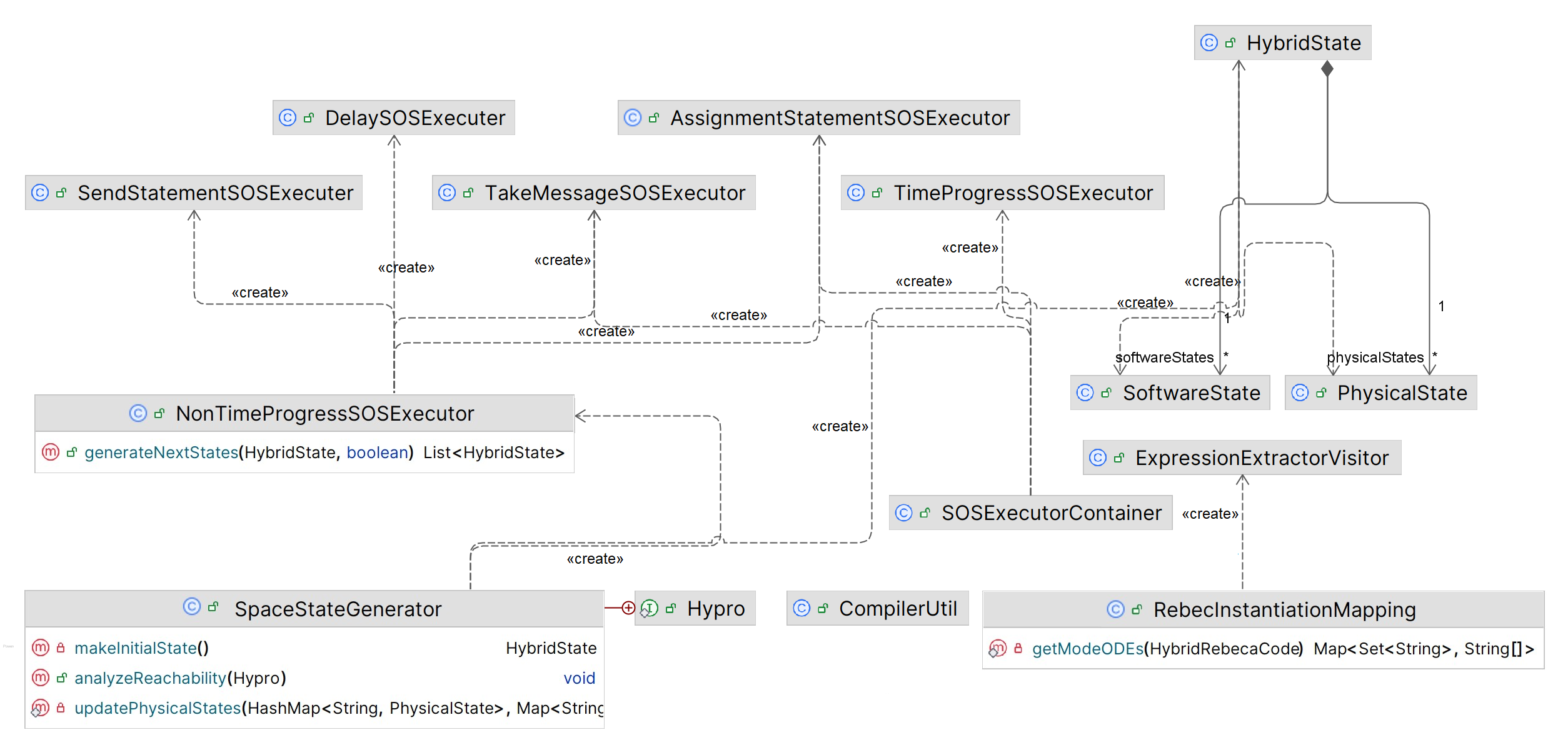}
	\caption{Class diagram}
	\label{Fig::Diagram}
\end{figure}

$\it SpaceStateGenerator$ class relies on a compiler, an extension of the Afra compiler\footnote{Avaiable at \url{https://github.com/rebeca-lang/org.rebecalang.compiler}}
called $\it CompilerUtil$, to convert a given Hybrid Rebeca model to its Abstract Syntax Tree (AST). 
The AST 
provides the necessary for defining the global states and the description of a mode of physical actors. 

The states of the semantic model are represented as instances of the class $\it HybridState$ which is a list of $\textit{PhysicalState}$ and $\textit{SoftwareState}$ classes, representing the continuous and discrete components of the model, respectively. The method \texttt{makeInitialState()} makes the initial state by extracting the instantiated actors from the AST and making the local states of actors based on their corresponding reactive class in AST. It executes the constructor for each instance in the order defined in the main by extracting the constructor's statements from AST. 
The constructors 
initialize the actor's state variables and message queues. 
Each actor's local state is defined by a $\textit{variableValuation}$ map. After initializing the reactive and physical rebecs in separate $\it HashMap$s, the \texttt{makeInitialState()} returns an instance of $\textit{HybridState}$. 

In our implementation, we use Java’s 
$\it LinkedList$ class to manage state exploration using a Breadth-First Search (BFS) approach initialized by our initial state. We dequeue a state $s$ and generate its next possible states by calling \texttt{generateNextStates($s$)},  a method of the class $\it NonTimeProgressSOSExecutor$. This class is responsible for applying non-time progressing SOS rules on the given state to generate the next possible states. 
We have dedicated classes of $\it SOSExecuter$ for each SOS rule. For instance for $\it Assign$ rule, we have the $\it SOSExecuterAssign$. Thus, $\it NonTimeProgressSOSExecutor$ aggregates a set of $\it SOSExecuter$s. 
These executors first check the applicability of the rule to the current state. If applicable, they generate the next possible states from the current one. 





After generating all possible next states without advancing time, 
the time progress SOS rule should be applied to advance these states forward in time. We compute the new global state based on $\it progressTime$ time and compute the flowpipes of real variables for each state returned by the method \texttt{generateNextStates}. 
To this aim, we collect ODEs for each physical rebec to call \( \textit{Hypro} \). As there are a number of instances from a Physical class, we should rewrite the specified ODE corresponding to its active mode by renaming the state variables to a fresh name. We append the name of each instance to the variables in ODEs. The specifications of ODEs are found from AST while the active mode of each instance is captured from its local state. For the sake of efficiency, we construct the ODEs of modes of each instance once by maintaining a mapping that associates the modes of each physical class with their corresponding ODEs. The \texttt{getModeODEs()} method of $\it RebecInstantiationMapping$ class initializes this mapping which is of type \texttt{Map<Set<String>, String[]>} in the variable $\it modeToODEs$, which maps each physical rebec and its mode with ODEs in that mode, captured as an array of String. This function exploits the $\it ExpressionExtractorVisitor$ class to handle the renaming. We use \emph{visitor pattern} to manage the rewriting for the hierarchy classes of ODE expressions. 
%
Generally speaking, we designated the class $\it RebecInstantiationMapping$, which serves as a centralized mapping between rebec instances and their corresponding types, known rebecs and $\it modeToODEs$
. These \( \textit{ODEs} \), along with the current \( \textit{intervals} \) of real variables and the time interval 
are passed to \textit{Hypro}. This tool computes the interval of each real variable. 

We update the physical rebecs with the computed values, obtained by invoking the method 
\texttt{updatePhysicalStates()}. This method evaluates each physical rebec's invariants and guards 
using the $\it ExpressionEvaluatorVisitor$, checks whether the guard conditions are satisfied, and if the rebec’s behavior adheres to its invariant constraints. We use \emph{visitor pattern} to handle the evaluation of the hierarchy classes of conditional expressions. Depending on the results of these checks, the state-space generator either continues to explore the current state or adjusts the physical state by enforcing the appropriate guard transitions. Then, the updated physical states are then re-enqueued for further exploration by calling \texttt{generateNextStates()}. State exploration proceeds until all reachable states have been generated or the simulation time exceeds the predefined upper bound.


\end{document}